\documentclass[aps,prd,showpacs,twocolumn,nofootinbib,superscriptaddress,amsmath,amssymb,8pt]{revtex4-2}
\usepackage{amsmath,amssymb,amsfonts}
\usepackage{bm}
\usepackage{hyperref}
\usepackage{slashed,braket}
\usepackage{color}
\usepackage{graphicx,graphics}
\usepackage[title,titletoc]{appendix}

\usepackage{tikz}\usetikzlibrary{shapes.geometric, arrows}\tikzstyle{method} = [rectangle, rounded corners, minimum width=3cm, minimum height=1cm, text centered, draw=black]\tikzstyle{obj} = [rectangle, minimum height=1cm, text centered, draw=black]\tikzstyle{arrow} = [thick,->,>=stealth]

\newcommand{\Tr}{\operatorname{Tr}}

\newcommand{\tr}{\operatorname{tr}}

\newcommand{\be}{\begin{eqnarray}}
\newcommand{\ee}{\end{eqnarray}}
\newcommand{\pd}{\partial}
\newcommand{\ep}{\epsilon}
\newcommand{\ra}{\to}
\newcommand{\tab}{\,}
\newcommand{\om}{\omega}

\newcommand{\mzz}[1]{{#1}}

{\color{blue}}%
{}

\begin{document}

\title{Weyl orbits as probe of chiral separation effect in  magnetic Weyl semimetals}

\author{M.A.Zubkov}
\email{zubkov.mike@gmail.com}
\affiliation{Physics Department, Ariel University, Ariel 40700, Israel}

\date{\today}

\begin{abstract}
We consider magnetic Weyl semimetals. First of all we review relation of intrinsic anomalous Hall conductivity, band contribution to intrinsic magnetic moment, and the conductivity of chiral separation effect (CSE) to the topological invariants written in terms of the Wigner transformed Green functions (with effects of interaction and disorder taken into account). Next, we concentrate on the CSE. The corresponding bulk axial current is accompanied by the flow of the states in momentum space along the Fermi arcs. Together with the bulk CSE current this flow forms closed Weyl orbits. Their detection can be considered as experimental discovery of chiral separation effect. Previously it was proposed to detect Weyl orbits through the observation of quantum oscillations \cite{Potter_2014} . We propose the alternative way to detect existence of Weyl orbits through the observation of their contributions to  Hall conductance.
\end{abstract}

\maketitle

\tableofcontents

\section{Introduction}

There are many analogies between condensed matter physics and physics of elementary particles. Both theories share the same formalism of quantum field theory. The important difference is that in condensed matter physics typically the basic symmetry of high energy physics - the Lorentz symmetry is absent. However, the discovery of topological semimetals (Dirac and Weyl semimetals) makes the two fields much more close to each other since in these materials there is an emergent relativistic symmetry \cite{semimetal_effects6,semimetal_effects10,semimetal_effects11,semimetal_effects12,semimetal_effects13,Zyuzin:2012tv,tewary,16}.
Therefore, these materials simulate in laboratory physics of elementary particles. Namely, the electronic quasiparticles are described by Dirac equation. Typically, the emergent vierbein is anisotropic, which reflects anisotropy of Fermi velocity. Apart from this the analogy is complete as long as we do not take into account interactions. The interactions break emergent Lorentz symmetry because the action of interaction carriers (phonons and photons traveling inside the substance) is not relativistic invariant.   Nevertheless,  these materials remain a unique laboratory for observation in laboratory of various high energy physics effects.

The important class of effects shared by solid state physics and physics of elementary particles is the class of non - dissipative transport effects  \cite{Metl,Kharzeev:2013ffa,Kharzeev:2015znc,Kharzeev:2009mf,ref:HIC,9,Landsteiner:2012kd,semimetal_effects6,Gorbar:2015wya,Miransky:2015ava,Valgushev:2015pjn,Buividovich:2015ara,Buividovich:2014dha,Buividovich:2013hza}.These effects are of topological nature and, therefore, their conductivities under certain conditions are robust to smooth modification of the system. This allows to calculate these conductivities in essentially complicatted systems with disorder and strong interactions. Recall that in Weyl semimetals even the Coulomb interactions become strong because of the renormalization of the fine structure constant in substance by electric permittivity.

One of the representatives of the family of non - dissipative transport effects is  chiral separation effect (CSE). It has been proposed by M.Metlitski and A.Zhitnitsky \cite{Metl}. It's essence is the appearance of axial current in the system of relativistic fermions directed along external magnetic field. This effect is expected to be observed in Weyl semimetals, with emergent relativistic invariance. In the system of conventional massless Dirac fermions the axial current is proportional to the external magnetic field strength $F_{ij}$ and to fermion chemical potential $\mu$:
\begin{equation}
	J_5^k = \frac{1}{4\pi^2}\epsilon^{ijk0} \mu F_{ij}\label{1}
\end{equation}

CSE is related intimately to chiral anomaly \cite{Zyuzin:2012tv}). The same refers to its cousin - the Chiral Magnetic Effect  (CME) \cite{Vilenkin,CME,Kharzeev:2013ffa,Kharzeev:2009pj,SonYamamoto2012}. Unlike the CSE \cite{Gorbar:2015wya} in thermal equilibrium there is no  CME  \cite{Valgushev:2015pjn,Buividovich:2015ara,Buividovich:2014dha,Buividovich:2013hza,Z2016_1,Z2016_2,nogo,nogo2,BLZ2021}. It comes back out of equilibrium \cite{BLZ2022}. According to the common lore the CME might be observed through the negative magnetoresistance of Weyl semimetals. The core idea is that in the presence of both electric and magnetic fields chiral anomaly pumps particles from "vacuum" (i.e. from the occupied bands of spectrum) \cite{Nielsen:1983rb}. These particles participate in electric current, which is to be seen as a contribution to conductivity  \cite{ZrTe5}.

There is a plenty of theoretical works on CSE. Analytical methods have been applied in \cite{KZ2017,SZ2020} to calculate the
CSE current in the lattice models. The results of numerical simulations have been reported in
 \cite{Buividovich:2013hza,Brandt2022AnomalousTP}. The general result of numerical results is that CSE conductivity is suppressed (compared to the naive result of non - interacting theory) in quark matter above the confinement - deconfinement crossover temperature.

Interaction corrections to CSE in quantum electrodynamics (QED) have been considered in \cite{Shovkovy}, where it was found that for the system of massive fermions the radiative corrections exist and depend on the infrared cutoff. In \cite{zubkov2023effect} the radiative corrections to the CSE current have been considered in QCD. It appears that the {\it renormalized} axial current does not receive corrections at zero temperature as long as the chiral symmetry is restored (this occurs, presumably, at sufficiently large baryon chemical potential). In the present paper we extend the derivation of \cite{zubkov2023effect} to CSE in Weyl semimetals, and demonstrate that the CSE conductivity is proportional to a topological invariant composed of the two - point electron Green function.

As in \cite{zubkov2023effect} we use Wigner - Weyl formalism in order to take into account the inhomogneities.
This calculus \cite{Weyl,Wigner} was originally invented by Groenewold and Moyal to non - relativistic quantum mechanics, where it replaced the operator formalism \cite{Groenewold,Moyal}.  It was used extensively also in quantum field theory, containing its applications to condensed matter physics. In our work we follow closely the version of this calculus reported in \cite{shitade,onoda1,onoda2,onoda3,onoda4,mokrousov}.

The essential feature of the Wigner - Weyl calculus used in the present paper is that it is adopted to the lattice models with compact Brillouin zone. It is worth mentioning that originally
Wigner - Weyl calculus was elaborated for  the continuous systems with infinite momentum space. The definition of its version for the lattice models was discussed in several works  \cite{Schwinger,Buot1,Buot2,Buot3,Wootters,Leonhardt,Kasperkovitz,Ligabo}. There are  problems specific for such a definition. The rigorous version of lattice Wigner - Weyl calculus for the infinte lattice has been proposed in \cite{FZ2019_2}. Its extension to the models defined on the finite lattice was discussed in \cite{Z2023}. In the present paper we rely on the simplified version of Wigner - Weyl calculus of \cite{FZ2019_2,Z2023}. This is the so - called approximate lattice Wigner - Weyl calculus - some of the basic properties of Weyl symbol of operator and Moyal product are approximate in this formalism, which is the price for the simplicity of formulation. For the details of this formalism see \cite{Zhang_Zubkov_PRD_2019,Suleymanov_Zubkov_2019,Fialk_Zubkov_2020_sym,Zhang_Zubkov_JETP_2019,Zhang_Zubkov_PhysLet_2020}. Very roughly, this calculus may be applied to the lattice systems, in which spatial  inhomogeneity is weak. In particular,  magnetic field should be much smaller than $10^5$ Tesla - which occurs always in real solid state systems, where maximal value of magnetic field does not exceed $100$ Tesla.

There were expectations that the CSE may be observed in the quark - gluon plasma (QGP) phase of quark matter, where confinement of quarks is absent and the chiral symmetry is restored. The quark gluon matter exists in the QGP phase within the fireballs that appear in heavy ion collisions   \cite{Kharzeev:2015znc,Kharzeev:2009mf,Kharzeev:2013ffa,ref:HIC}. In the non - central collisions the fireballs experience strong magnetic field \cite{QCDphases,1,2,3,4,5,6,7,8,9,10}.
After the decay of the fireball, the signature of the CSE might be observed through the asymmetry of outgoing particles. However, in practise the CSE gives subdominant contribution to the corresponding observed quantities, while the dominant contribution is due to the chiral vortical effect (CVE). As a result the CSE has not been observed so far in particle physics.

In Weyl semimetals also the experimental observation of the CSE has not been reported until now. Even it has not been proposed clearly how it can be observed in principle. The reason for this is that the majority of transport phenomena are related typically to charge transport, energy transport, spin transport, etc. The CSE neither transfer electric charge nor energy, nor spin. It transfers chirality, but it is not clear ad hoc how to identify chirality at the boundaries of the sample. 

First of all, let us discuss the very definition of the axial current in Weyl semimetal. It is defined as the electric current carried by the right – handed quasiparticles minus electric current carried by the left – handed quasiparticles. In the absence of interactions and at zero temperature the right – handed quasiparticles are defined as those that reside in momentum space in a vicinity of the right – handed Weyl point (the one with positive value of the topological invariant responsible for stability of Fermi points – see below Eq. (\ref{N3})). Correspondingly, the left – handed quasiparticles are those, which reside in the vicinity of the left – handed Weyl point. For small enough chemical potential these definitions are non – ambiguous because the right – handed Weyl particles occupy the small vicinity of momentum space around the Weyl point bounded by the corresponding piece of the Fermi surface. Correspondingly, now the axial current is a well – defined quantity provided that temperature is zero and interactions are absent. The presence of finite temperature makes the pattern ambiguous because the temporal quasiparticles do not necessarily reside in small vicinities of the Weyl points. However, {\it we restrict ourselves by very small temperatures}, when this effect may be neglected. The presence of interactions also changes the pattern – now, strictly speaking, the Fermi surface does not enclose positions of the quasiparticles. However, the very concept of Fermi surface remains alive as the position of the pole of the complete interacting Green function. In particular, Weyl point is still defined as the position of Fermi surface when it is reduced to a point at a certain value of chemical potential $\mu$. Then at any value of $\mu$ the chirality operator $\gamma_5$ is defined as a function in phase space.  This is function in momentum space, which equals $+1$ in a small vicinity of the right – handed Weyl point and equals $-1$  in a small vicinity of the left – handed Weyl point. (Notice that the position of the Weyl point may depend on coordinates, and therefore the chirality operator depends on coordinates as well.) It is assumed that the considered values of chemical potential are such that the Fermi pockets remain within the mentioned vicinities of the Weyl points. Then our definition of renormalized axial current (see below Eq. (\ref{J5R})) defines non – ambiguously the axial current in Weyl semimetal also in the presence of interactions (provided that the effects of nonzero temperature are disregarded).

The present paper is devoted to the possible ways to observe CSE. Our propositions are related to phenomena that connect physics of the boundary with physics in the bulk through the so - called Weyl orbits. The existence of Weyl orbits has been proposed for the first time in \cite{Potter_2014}. The Weyl orbit consists of motion of a right - handed quasiparticle in the bulk along external magnetic field, sliding on the surface of the sample along the Fermi arc that connects the Weyl points of opposite chirality, back motion of the left - handed quasiparticle inside the bulk in direction opposite to that of magnetic field, and again sliding along the Fermi arc, which completes the circle. The pieces of the Weyl orbit directed along magnetic field realize the bulk axial CSE current. Energy of a particle, which moves along the Weyl orbit in phase space is quantized, and it was proposed in \cite{Potter_2014} to observe these quantized energies through the frequencies of quantum oscillations (for example, of density of states).
Interaction effect on the mentioned frequencies has been considered in \cite{gorbar2014quantum}. In  \cite{zhang2021cycling} several experiments were reviewed, and results of some of them may, possibly, be interpreted as observation of the oscillations proposed in \cite{Potter_2014}.
In particular, in \cite{moll2016transport} the observation of the Shubnikov-de Haas oscillations in Focused Ion Beam prepared microstructures of Cd$_3$As$_2$ were reported that are claimed to be consistent with the theoretically predicted Weyl
orbits. The typical thickness of these microstructures was of the order of $150$ nm, and then at the temperatures below $100$ K typically one frequency of the oscillations appears, which is consistent with ARPES and STM experiments. The main result of \cite{moll2016transport} is that in the presence of magnetic field perpendicular to a certain  plane of the crystal lattice the new frequency appears. This new frequency, according to the authors of \cite{moll2016transport} is  consistent with the existence of Weyl orbits. The authors of \cite{moll2016transport} also investigated the dependence of the observed frequency on the angle between the magnetic field and the planes of the crystal and on the sample thickness. The further experimental investigation of the oscillation frequencies and their relations to the Weyl orbits has been performed with similar results in \cite{zhang2017evolution,zheng2017recognition,zhang2019ultrahigh,zhang2019quantum}. In principle, the observation of the mentioned frequencies may serve as the indirect way to detect experimentally the CSE in Weyl/Dirac semimetals. The main potential drawback of this statement is that we cannot exclude at the present moment that another source of the given oscillation frequency may exist, not related at all to Weyl orbits.      

We propose the alternative way to detect the Weyl orbits, and, therefore, to observe the CSE.  Namely, we will consider the sample of magnetic Weyl semimetal in the presence of external magnetic field, and in addition, in the presence of voltage between the sides of the sample. It appears that the specific contribution to Hall conductance appears, which carries information about the bulk CSE. The actual experimental setup includes application of electric current and measuring Hall voltage between the sides of the sample. As well as the method based on the observation of quantum oscillations this is also an indirect way to observe the CSE since it goes through the detection of the Weyl orbits rather than the bulk CSE current itself. It is worth mentioning that relation between Weyl orbits and Hall effect has been considered also in another context for Dirac (rather than Weyl) semimetals in \cite{zhang2017evolution,zhang2019quantum}.

The paper is organized as follows. In Sect. \ref{basis}  we discuss  topological conductivities in Weyl semimetals. Namely, the Wigner - Weyl calculus is applied to the derivation of intrinsic contribution to anomalous quantum Hall effect and to the conductivity of chiral separation effect in Weyl semimetals. In principle, the results presented in this section are known. However, we present them here in such a form that allows us to take into account both interactions and disorder (which, to the best of our knowledge has not been done before).

 In Sect. \ref{magnWeyl} we apply results of the previous section to magnetic Weyl semimetals. Namely, we calculate contribution of the Weyl points to intrinsic magnetic moment, and their contribution to anomalous Hall effect. The whole pattern of the latter effect has been discussed here for completeness.

 In Sect. \ref{WeylOrbit} we describe Weyl orbits and their relation to chiral separation effect. Also in this section we reproduce seminal results of \cite{Potter_2014} on the quantization of the Weyl orbit energy.

 In Sect. \ref{prop} we describe the alternative way to observe the Weyl orbits and the chiral separation effect based on the Hall effect.

 In Sect. \ref{concl} we end with the conclusions, and list the obtained results.

\section{Topological conductivities in Weyl semimetals}
\label{basis}

\begin{figure}[h]
	\centering  %
	\includegraphics[width=0.9\linewidth]{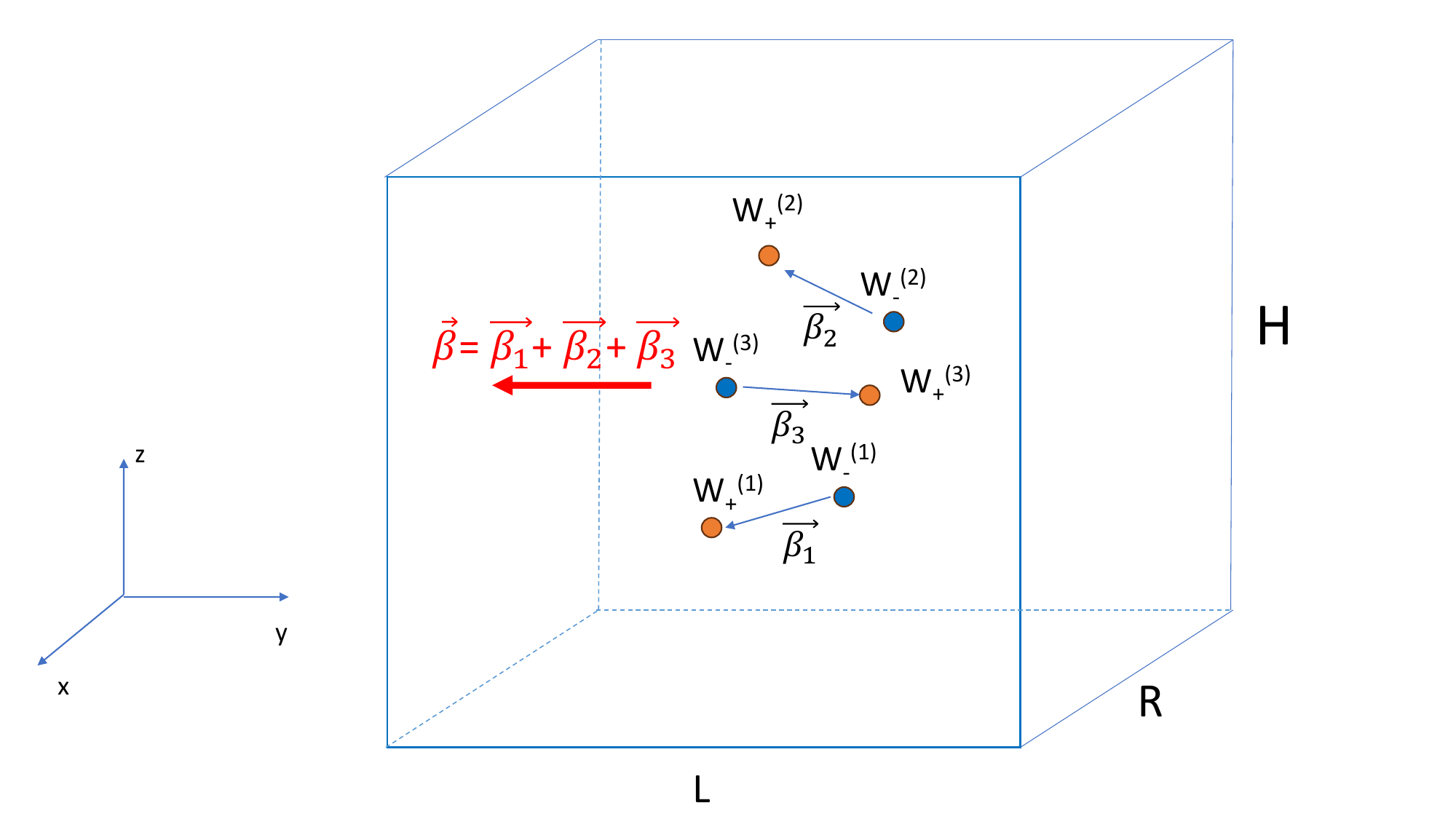}  %
	\caption{The sample of Weyl semimetal of hight $H$, width $R$ and length $L$. Inside the sample the three pairs of Weyl points are drawn schematically. We denote by $W^{(i)}_+$ the $i$ - th right - handed Weyl point, and by $W^{(i)}_-$ the $i$ - th left handed Weyl point. Vector connecting the Weyl point in the $i$ - th pair is denoted by $\beta_i$. The sum of these vectors results in the overall vector $\beta$, which is nonzero for magnetic Weyl semimetals.  }  %
	\label{fig0}   %
\end{figure}

\subsection{Preliminaries}
	
 In Weyl semimetals the electron energy bands cross each other forming Weyl points. In a small vicinity of each Weyl point the quasiparticle Hamiltonian reads
 \begin{equation}
 	{\cal H}(\vec{p})  = \pm v_a^i \sigma^a p_i + f^i p_i \label{Ham}
 \end{equation}
where $\vec{p}$ is momentum while tensor $v_a^i$ represents the anisotropic Fermi velocity. In real materials this tensor is always nontrivial. In the next sections of the present paper for simplicity we will consider the isotropic case with $v_a^i = v_F \delta_a^i$. However, in the present section the particular form of tensor $v_a^i$ is irrelevant. Vector $\vec{f}$ is responsible for the tilting of Dirac cone.  Sign $+$ is chosen in Eq. (\ref{Ham}) for the right - handed Weyl point, while sign minus is  chosen for the left - handed Weyl point. These points always come in pairs. In the ordinary type I Weyl semimetals (with time reversal symmetry) vector $\vec{f}$ vanishes, and the Dirac cone is not tilted. For the type I Weyl semimetals with broken time reversal symmetry vector $\vec{f}$ may be nonzero, but the Weyl points remain Fermi points. In the so - called type II Weyl semimetals the Dirac cone is overtilted, and in the vicinity of each Weyl point the Fermi pocket appears. This occurs if
$$
\|(v^{-1})^a_i f^i \|>1, \quad (v^{-1})^a_i v^i_b = \delta^a_b
$$
The appearance of this type of Weyl point has been proposed for the first time in \cite{VZ2014}. Later the existence of such Weyl points has been predicted in real materials \cite{soluyanov2015type} (WTe$_2$). In the present paper we will not consider the type II Weyl semimetals and concentrate on ordinary type I Weyl semimetals.

In Dirac semimetals (Cd$_3$As$_2$, Na$_3$Bi, etc) the positions of the two Weyl points in each pair coincide, and we are speaking of Dirac points. In Weyl semimetals in each pair the positions of Weyl points in momentum space differ by vector $\vec{\beta}_k$ (index $k$ enumerates the pairs). In the non - magnetic Weyl semimetals (for example, TaAs) the sum $\sum_k \vec{\beta}_k$ vanishes. In magnetic Weyl semimetals (for example, Co$_3$Sn$_2$S$_2$) vector
$$
\vec{\beta} = \sum_k \vec{\beta}_k
$$
is nonzero, which results in the nonzero contribution to the total magnetic moment of the sample (see below Sect. \ref{Magnetic}). However, typically, the magnetization of the sample is dominated by the other mechanisms, and the contribution of band topology to the total magnetization is sub - dominant.

The sample of typical Weyl semimetal is  represented schematically on Fig. \ref{fig0}. This sample contains $3$ pairs of Weyl points (as Co$_3$Sn$_2$S$_2$). The hight of the sample (its linear size in direction of axis $z$) is denoted by $H$, the width (the size in direction of axis $x$) is denoted by $R$, the length (the size in direction of axis $y$) is denoted by $L$. In the following we will have in mind such a sample of rectangular form during our considerations of phenomena related to the chiral separation effect (CSE).

\subsection{Partition function of non - interacting electrons}

In the next sections we will use Wigner - Weyl calculus in order to calculate axial current of chiral separation effect,  magnetization and Hall conductivity of Weyl semimetals. Here we present brief description of this formalism based on our previous publications \cite{ZZ2019_0,ZZ2019_1,ZZ2021,ZW2019,SZZ2020}. We consider the fermion system defined on the crystal lattice. First we neglect interactions.

Partition function is expressed through the inverse bare Green function. We refer to it as to Dirac operator, which is denoted by  $\hat Q$. In the absence of interactions it is expressed through one - particle Hamiltonian $\cal H$ as ${\hat Q} = i \omega - \hat{\cal H}$. The partition function  may be represented in the form of functional integral over Grassmann - valued  fields $\psi, \bar{\psi}$:
\be
Z = \int D\bar{\psi }D\psi
\,\, e^{S[\psi ,\bar\psi  ]}
\label{Z01}
\ee
Here $S$ is the action, which may be written in momentum space as
\begin{equation}\begin{aligned}
		S[\psi ,\bar\psi  ]&=\int_{\mathcal M} \frac{d^D{p}}{|\mathcal M|}\bar\psi({p}) \hat Q(i\partial_{p},{p})\psi({p})\\
		&=\int_{\mathcal M} \frac{d^D{p}}{|\mathcal M|}\bar\psi^a({p}) \hat Q^{ab}(i\partial_{p},{p})\psi^b({p})
\end{aligned}  \end{equation}
In the last expression $D=4$ is the dimensionality of space - time, while Weyl symbol of operator is used:
\begin{equation} \begin{aligned}
		{Q}^{}_W(x,p) \equiv \int_{\cal M} dq e^{ix q} \langle {p+q/2}| \hat{Q} | {p-q/2}\rangle\label{Q_W}
	\end{aligned}\,,
\end{equation}
Here $\ket{p}$ is the eigenvector of lattice momentum,
while
\be W^{ba}(x,{p})=\int_{\cal M} dq e^{ix q} \psi^b (p+q/2) \bar{\psi}^a (p-q/2)  \ee
may be considered as Weyl symbol of an operator  with matrix elements
${\psi(p_1)}{\bar{\psi}}(p_2)$, which are Grassmann rather than ordinary numbers. 

 In order to unify space and imaginary time notations we discretize both space coordinates and imaginary time. By $\cal M$ we denote momentum space, while $|{\cal M}|$ is its volume. The discretization of imaginary time may be taken off, which brings us back to   $\hat{Q} = i \omega - \hat{\cal H}$.
 
 When the system lagrangian is not changed drastically at the distance of lattice spacing, the external electromagnetic potential is taken into account through 
\textit{Peierls} substitution.  If in the absence of external electromagnetic field $ A$ function $Q_W$ had the form $Q^{(0)}_W(r,{p}) \equiv Q^{A=0}_W(r,{p})$, while operator $\hat{Q}$ had the form $\hat{Q}^{(0)}(i\partial_p,p) \equiv Q^{A=0}(i\partial_p,p)$, then in the presence of $ A$ the Weyl symbol of $\hat{Q}$ receives the form  
\be Q_W(r,{p}) = Q^{(0)}_W(r,{p}-A(r))\equiv Q^{A=0}_W(r,{p}-A(r))\label{QWA} \ee
This quantity depends on coordinates $r$ both through the dependence of electromagnetic field on coordinates and because of the inhomogeneity of another type. At the same time the Dirac operator itself becomes
 \be \hat{Q} = \hat{Q}^{(0)}(i\partial_p,p-A(i\partial_p))\equiv \hat{Q}^{(A=0)}(i\partial_p,p-A(i\partial_p))\label{QA} \ee
Electron Green function in the presence of external gauge field is defined as
\begin{equation}\begin{aligned}
		&G({p}_1,{p}_2)=\bra{{p}_1} \hat{G} \ket{{p}_2}\\&=
		\frac{1}{Z}\int D\bar\psi D\psi \bar\psi({p}_2) \psi({p}_1) \\&\exp\left(\int \frac{d^D{p}}{|\mathcal M|}\bar\psi({p})  \hat{Q}^{(0)}(i\partial_p,p-A(i\partial_p))\psi({p})\right)
\end{aligned}  \end{equation}

\subsection{Partition function in the presence of interactions}	

There are a lot of various interactions between electrons in solids. Those include exchange by phonons as well as Coulomb interactions. The latter are strong due to the expression for effective fine structure constant $\frac{e^2}{4 \pi \epsilon_0 \hbar v_F \epsilon}$ (here Fermi velocity $v_F$ is of the order of $10^{-3}$ speed of light, while dielectric permittivity $\epsilon$ is not large). Besides, we introduce the external electromagnetic field $A$. For definiteness we consider the interaction of electrons with the  fields $\Phi^{ab}$, when the interaction term in Lagrangian has the simplest form $L_I = \bar{\psi}^a \Phi_{ab} \psi^b$. Such a lagrangian is responsible, for example, for the description of exchange by optical phonons and, most importantly, for the description of inter - electron Coulomb interactions. This brings the partition function to the form	
\begin{equation} \begin{aligned}
		Z&=\int D\bar\psi D\psi D\Phi e^{-S_\Phi[\Phi]} \\&\exp
		\left(\int_{\mathcal M} \frac{d^D{p}}{|\mathcal M|}\bar\psi^a({p}) \hat{Q}^{(0)}(i\partial_p,p-A(i\partial_p))\psi^b({p})\right)
		\\&\exp
		\left(\int_{\mathcal M} \frac{d^D{p}d^Dq}{|\mathcal M|^2}\bar\psi^a({p+q}) \Phi_{ab}(q) \psi^b({p})\right)
\end{aligned} \end{equation}
 $S_\Phi[\Phi]$ is the action for $\Phi$. Both Dirac operator and the Green function depend explicitly on $ A$. While dependence of $Q_W$ on $ A$ is relatively simple: it is given by Paierls substitution, the dependence of $G_W$ on $ A$ is much more complicated.

Let us consider variation of partition function due to the variation of Dirac operator $\hat{Q}$. We will be interested further in the variations of $Q_W$ that are caused by the introduction of infinitely small external electromagnetic potential $\delta A(r)$. Its introduction is implemented as $\delta Q_W(r_n,q) = Q^{(0)}_W(r_n,q-A(r_n)-\delta A(r_n)) - Q^{(0)}_W(r_n,q-A(r_n))$. 
Then we obtain 
\begin{widetext}
\begin{equation} \begin{aligned}
		\delta \log Z&=-\frac{1}{Z}
		\int D\bar\psi D\psi D\Phi e^{-S_\Phi[\Phi]}
		\Big[\sum_{{x}_n} \int \frac{d^D{q}}{|\mathcal M|}
		\delta Q^{ab}_W({x}_n,q) W^{ba}({x}_n,{q})\Big]
		\\&\exp
		\left(\int_{\mathcal M} \frac{d^D{p}}{|\mathcal M|}\bar\psi^a({p}) \hat{Q}^{(0)}(i\partial_p,p-A(i\partial_p))\psi^b({p})\right)
		\exp
		\left(\int_{\mathcal M} \frac{d^D{p}d^Dk}{|\mathcal M|^2}\bar\psi^a({p+k}) \Phi_{ab}(k) \psi^b({p})\right)=\\
		&=\sum_{{r}_n} \int \frac{d^D{p}}{|\mathcal M|}
		\tr \left[ \delta Q_W({r}_n,{p}) {\bf G}_W({r}_n,{p}) \right]
\end{aligned} \end{equation}
\end{widetext}
Here by $\bf G$ we denote the complete interacting Green function. {\it One can see that the effect of interactions on the variation of partition function is reduced to the appearance of the interacting Green function istead of the non - interacting one.} When dependence of Dirac operator on coordinates is weak, i.e. may be neglected at the distance of the order of lattice spacing,  we replace the sum by an integral and obtain
\begin{equation} \begin{aligned}
		\delta \log Z&= \int d^D x  \int \frac{d^D{p}}{{ v}|\mathcal M|}
		\tr \left[ \delta Q_W(x,{p}) {\bf G}_W(x,{p}) \right]\\
		&= \int d^D x  \int \frac{d^D{p}}{(2\pi)^D}
		\tr \left[ \delta Q_W(x,{p}) {\bf G}_W(x,{p}) \right]
		\label{dlogZ}\end{aligned} \end{equation}
By $v$ we denote the elementary lattice cell volume. We use relation between momentum space volume $\mathcal M$ and the lattice cell volume  ${v} |{\mathcal M}| = (2\pi)^D$. Notice that the interacting Green function ${\bf G}_W(x,{p})$ is taken in the presence of external gauge field and depends on it.

We rewrite Eq. (\ref{dlogZ}) as
\begin{equation} \begin{aligned}
		&\delta \log Z= \tr \left[\hat {\bf G} \delta \hat Q  \right]=\Tr[{\bf G}_W\star \delta Q_W]=
		\Tr[ {\bf G}_W\delta Q_W]
\end{aligned} \end{equation}
where the Moyal product is
$$
\star = e^{\frac{i}{2} \left( \overleftarrow{\partial}_{x}\overrightarrow{\partial_p}-\overleftarrow{\partial_p}\overrightarrow{\partial}_{x}\right )}
$$
In these expressions by  $\bf G$ we denote the complete interacting two - point quark Green function while $G$ is the Green function in the presence of external field $\Phi$. The star may be removed here if $\delta Q_W(p,x)$ as a function of $x$ is localized in finite region of space.

Electric current appears as response of partition function to variation of external electromagnetic potential. The corresponding variation of $Q_W$ is given by 
\begin{eqnarray}
	&& \delta Q_W(x,p)=\pd_{A_i}Q^{(0)}_W(x,p-A)\delta A_i
\end{eqnarray}
which results in
\be
\delta Q_W=\pd_{A_i} Q_W \delta A_i=-\pd_{p_i} Q_W \delta A_i
\ee
Therefore, the electric current is expressed as
\be
j_i(x)=\frac{\delta \log Z}{\delta A_k(x)}=
-\int_{(2\pi)^D} \frac{d^Dp}{|\mathcal M|}
\tr \left[ {\bf G}_W(x,p) \partial_{p_i} Q_W(x,p)  \right]
\label{j_i}
\ee
Based on analogy with electric current the naive expression for local axial current density may be defined as
\be
j^5_k(x)= -
\int_{\mathcal M} \frac{d^Dp}{(2\pi)^D}
\tr \left[ \gamma^5(p,x)  {\bf G}_W(x,p) \pd_{p_k} Q_W(x,p)  \right]
\ee
Here $\gamma_5(p,x)$ is chirality operator, which is equal to $+1$ in a small vicinity (in momentum space) of right - handed Weyl point and is equal to $-1$ in the small vicinity of the left - handed Weyl point (position of Weyl point in general case depends on $x$). {\it By Weyl point position we understand here the pole of ${\bf G}_W(p,x)$ as a function of $p$ (when chemical potential is adjusted in such a way that the Fermi surface is reduced to a point). Therefore, this is the coordinate dependent Weyl point. As a result $\gamma_5(p,x)$ is real - valued function in phase space, which equals to $+1$ at the position $p_0(x)$ of the pole of ${\bf G}_W(p,x)$, where
	$$
{\bf G}^{-1}_W(p,x) \approx v_a^i(x) \sigma^a (p - p_0(x)) + f^i(x) (p - p_0(x))
	$$
	with certain functions $v$ and $f$.
At the positions of the left - handed Weyl point the value of $\gamma_5(p,x)$ equals $-1$. There the Wigner transformed Green function has an expansion
	$$
{\bf G}^{-1}_W(p,x) \approx -v_a^i(x) \sigma^a (p - p_0(x)) + f^i(x) (p - p_0(x))
$$
with the other functions $f$ and $v$.
In the intermediate region of phase space the value of $\gamma_5(p,x)$ interpolates between $+1$ and $-1$.
}

It is worth mentioning that the above given linear dependence of ${\bf G}^{-1}_W(p,x)$ on momentum corresponds to the value of the topological invariant ${\cal N}_3$ equal to $\pm 1$ (see below Eq. (\ref{N3})). In general case this invariant can have any integer values. However, the corresponding Weyl points are not stable. Without an additional symmetry, which forces the value of ${\cal N}_3$ to have, say, a value $+2$, any smooth modification of the system splits such a single Weyl point into the two Weyl points with minimal values ${\cal N}_3 = 1$ (see \cite{volovik2003}). For this reason we restrict in the present paper our consideration to the case of the Weyl points with minimal values of ${\cal N}_3$. \mzz{In this case the dispersion near Weyl point is linear, while more complicated dependense of energy on momentum would correspond to the larger values of ${\cal N}_3$ (see \cite{Volovik2003}).} It is worth mentioning, however, that there exists typically the infrared cutoff $\Lambda$ in the theory, and the model with the topologically protected Weyl points breaks down at momenta extremely close to the Weyl point. There the dependence of Hamiltonian on momentum indeed deviates from linear. However, this region of momenta does not contribute to the considered physical quantities at the typical values of Fermi level (counted from the level of Weyl points) above $v_F \Lambda$, where $v_F$ is Fermi velocity.

\subsection{Renormalized velocity and renormalized currents}
\label{SectRenorm}
{Expression $-\partial_{p_i} Q_W(x,p)$ represents  matrix of bare quasiparticle velocity. The electric current is  given by averaging of this velocity. In the presence of interactions the renormalized electron velocity is to be substituted to expression for electric current instead of the bare velocity. We denote by $\bf Q$ an operator inverse to $\bf G$, which is the Green function of electrons, in which contribution of electron self - energy is taken into account.  The renormalized velocity operator (better to say, its Weyl symbol) is given by
	\begin{equation}
		v_R=-\partial_{p_i} {\bf Q}_W(x,p)
	\end{equation}	
	It can be shown \cite{ZZ2022} that we can take the "renormalized" expression for the electric current density
	$$
	{\bf j}_k(x) =
	- \int_{\mathcal M}  \frac{d^Dp}{(2\pi)^D}
	\tr \left[ {\bf G}_W(x,p) \partial_{p_i} {\bf Q}_W(x,p)  \right]
	$$
	{\it Notice that in this expression the value of electron electric charge $e$ is not included. Besides, here and below we work in the system of units with $c = \hbar = 1$. In expressions that are intended to be compared to experiment we will restore the conventional definition of electric current (with $e$ included) and pass to the SI units. This will be pointed out explicitly after the corresponding expressions.}
	
	Based on an analogy to electric conductivity in \cite{zubkov2023effect} it has been proposed to calculate the renormalized axial current using  operator of renormalized velocity in place of the bare velocity:
		\be
		{\bf j}^5_k(x)= -
		\int_{\mathcal M} \frac{d^Dp}{(2\pi)^D}
		\tr \left[ \gamma^5(p,x)  {\bf G}_W(x,p) \pd_{p_k} {\bf Q}_W(x,p)  \right]\label{J5R}
		\ee
		It appears that for the calculation of CSE conductivity one should integrate in momentum space along the infinitely small hypersurface surrounding the position of Fermi surface/Fermi point.

	\subsection{Topological expressions for the intrinsic anomalous QHE conductivity and for the CSE conductivity}
	\label{QHECSE}
In Appendix \ref{AppendixB} using Wigner - Weyl calculus we derive  expressions for the intrinsic anomalous contribution to the QHE conductivity of magnetic Weyl semimetals and to the so - called CSE conductivity. Both conductivities are calculated in the limit of small temperature. Here we represent the corresponding results.
 
	\subsubsection{QHE conductivity}
	\label{QHE}
	The local renormalized electric current density 
	 averaged over the whole system volume $V$ can be calculated as 
	\be
	&&\bar{J}_i\equiv \frac{1}{\beta {V}}\sum_x {\bf j}_i(x)\nonumber\\&&=
	-\frac{1}{\beta {V}}\int d^Dx \int_{\mathcal M} \frac{d^Dp}{{v}|\mathcal M|}
	\tr \left[ {\bf G}_W(x,p) \partial_{p_i} {\bf Q}_W(x,p)  \right]
	\label{Ii}\ee
	Here $v$ is volume of the lattice cell. Response of this averaged current to external electric field gives the following expression for the (averaged) Hall conductivity
	 \cite{ZW2019}:
	\begin{equation}
		\sigma^H_{ij} = \frac{e^2}{h}\, \frac{\epsilon_{ijk}{\cal M}_k}{2\pi}\label{sigmaH}
	\end{equation}
	with
	\begin{eqnarray}
		{\cal M}_i &=&   \frac{1}{3! \, 4\pi^2\,  V} \epsilon_{ijkl} \int d^3 r\, {d^4 p}\, {\rm tr}\,{\bf G}_W^{(0)} \star \partial_{p_j} {\bf Q}_W^{(0)} \nonumber\\&& \star {\bf G}_W^{(0)} \star \partial_{p_k} {\bf Q}_W^{(0)} \star {\bf G}_W^{(0)} \star \partial_{p_l} {\bf Q}^{(0)}_W
		\label{M00}
	\end{eqnarray}
The last expression would be topological invariant for the topological insulators. However, the topological invariance is lost for Weyl semimetals. It remains partially: the value of $\cal M$ is robust to those modifications of the system, which do not alter the values of $\bf G$ at their poles (i.e. at the positions of Weyl points). It is worth mentioning that the obtained expression is valid for vanishing temperature.

Notice that even at very small temperature in addition to the intrinsic contribution (given by Eq. (\ref{sigmaH})) to anomalous Hall conductivity in Weyl semimetal there are the non - topological contributions. Those contributions are related to various scattering processes. For more details see below Sect. \ref{AQHE}.

\subsubsection{CSE conductivity}
	
	In a similar way we can consider the renormalized axial current density averaged over the whole system volume:
	\be
	&&\bar{J}_i^5\equiv \frac{1}{\beta {V}}\sum_x {\bf j}_i^5(x)\nonumber\\&&=
	-\frac{1}{\beta {V}}\int d^Dx \int_{\mathcal M} \frac{d^Dp}{{v}|\mathcal M|}
	\tr \left[\gamma^5(p,x)  {\bf G}_W(x,p) \partial_{p_i} {\bf Q}_W(x,p)  \right]
	\label{Ii5}\ee
	Its response to external magnetic field strength $B$ and to the variation of chemical potential $\delta \mu$ may be represented as
		$$
	\bar{J}_k^5(x)=\mathcal{\sigma}_{CSE} B_{k}\delta\mu
	$$
with the CSE conductivity
	\begin{equation}
		\sigma_{CSE} = \frac{\mathcal{N}}{2\pi^2}\label{sigmaH}
	\end{equation}
	and
	\begin{widetext}
		\begin{eqnarray}
			\mathcal{N}
			&=&\frac{1}{48 \pi^2 {V}}
			\int d^3x \int_{\Sigma(x)}
					\tr \Bigg[\gamma^5(p,x)
			{\bf G}_W^{(0)}\star d {\bf Q}_W^{(0)} \star {\bf G}_W^{(0)}
			\wedge \star d {\bf Q}_W^{(0)}\star {\bf G}_W^{(0)} \star \wedge d {\bf Q}_W^{(0)}
			\Bigg]\label{N0}
		\end{eqnarray}
	\end{widetext}
Here integration in momentum space is over the hypersurface $\Sigma(x)$ that surrounds the positions of the Weyl points.
	
	In the particular case, when background is homogeneous (no elastic deformations, impurities, etc), we obtain:
	\begin{widetext}
		\begin{eqnarray}
			\mathcal{N}&=&\frac{1}{48 \pi^2 }
			\int_{\Sigma_3}
			\tr \Bigg[\gamma^5(p,x)
			{\bf G}_W^{(0)} d {\bf Q}_W^{(0)}  {\bf G}_W^{(0)}
			\wedge  d {\bf Q}_W^{(0)} {\bf G}_W^{(0)}  \wedge d {\bf Q}_W^{(0)}
			\Bigg]
		\end{eqnarray}
	\end{widetext}
	where  $\bf Q = {\bf G}^{-1}$. Index $^{(0)}$ points out that external electromagnetic field is set to zero, while chemical potential is set to the level of the Weyl points. This level  is assumed to be equal for all Weyl points. This assumption is not always fulfilled, and then the Fermi points are transformed to the Fermi pockets. We assume, however, that those Fermi pockets remain sufficiently small, so that the value of $\gamma_5(p,x)$ can be taken constant along them. Then the above expressions remain valid. However, surface $\Sigma(x)$ surrounds the position of the Fermi pockets.

	In practise in type I Weyl semimetals the value of $\mathcal N$ equals to the number of the pairs of Weyl fermions. Thus we prove here that the interactions do not modify expression for the CSE conductivity.

\section{Properties of magnetic Weyl semimetals}
\label{magnWeyl}	
\subsection{Magnetic moment and topological invariant responsible for the Weyl points}

\label{Magnetic}

We can define the magnetic moment of the system of conducting electrons as response of the thermodynamical potential to external magnetic field.  First we calculate response of the partition function to chemical potential:
\begin{eqnarray}
&&	\delta \, {\rm log}\, Z  = -i \int d^4 r\, \frac{d^4 p}{(2\pi)^4}\, {\rm tr}\, {\bf G}_W(p,r)\partial_{p_4} Q_W(p,r)\,\delta \mu \nonumber\\  & = &-i \int d^4 r\, \frac{d^4 p}{(2\pi)^4}\, {\rm tr}\, {\bf G}_W(p,r)\star \partial_{p_4} Q_W(p,r)\,\delta \mu \label{J4}
\end{eqnarray}
In the second line of the above expression we introduce the star product, which is possible in the absence of external electric field.

Next, using the same machinery as in \cite{ZZ2022} we are able to prove that this expression equals to the one, in which "bare" Dirac operator is replaced by the renormalized one. The only difference is the derivative with respect to momentum. In \cite{ZZ2022} this derivative is with respect to the spatial components of momentum while in the present paper the derivative is with respect to the fourth component of momentum, i.e. with respect to the Matsubara frequency. (Therefore, we are forced to consider vanishing temperature, when the sum over Matsubara frequencies is reduced to an integral.) Besides, in \cite{ZZ2022} the dimension was equal to $2+1 = 3$, while here it is $3+1=4$. Apart from this the algebra represented in Sect. 6 of \cite{ZZ2022} is to be applied directly to Eq. (\ref{J4}). The proof works {\it to all orders of perturbation theory}, and we arrive at:
\begin{eqnarray}
	\delta \, {\rm log}\, Z  & = &- i \int d^4 r\, \frac{d^4 p}{(2\pi)^4}\, {\rm tr}\, {\bf G}_W\star \partial_{p_4} {\bf Q}_W\,\delta \mu \label{J4_}
\end{eqnarray}
Next, we calculate the response to the external field strength $\delta F_{ij}= \epsilon_{ijk} B_k$ using expansion of
Eqs. (\ref{exp}), (\ref{exp1}). This results in
\begin{widetext}
\begin{eqnarray}
	\delta \, {\rm log}\, Z
	&=& \frac{1}{2}\epsilon_{ijk} \int d^4 r\, \frac{d^4 p}{(2\pi)^4}\, {\rm tr}\,{\bf G}_W^{(0)} \star \partial_{p_i} {\bf Q}_W^{(0)} \star {\bf G}_W^{(0)} \star \partial_{p_j} {\bf Q}_W^{(0)} \star {\bf G}_W^{(0)}\star \partial_{p_4} {\bf Q}^{(0)}_W(p,r)\,\delta \mu  \, \delta B_{k}
\end{eqnarray}
\end{widetext} 	
Recall that the integration over imaginary time $r^4$ goes from $0$ to inverse temperature $1/T$.	
For the system of electric currents in the presence of external magnetic field $\vec{B}$ (the one, which is not created by the currents themselves) the response of the thermodynamical potential to variation of magnetic field is given by 
$$
\delta \Omega = \vec{M} \delta \vec{B}
$$
Relation between the thermodynamical potential $\Omega$ and partition function ${\rm log}\, Z = - \Omega/T$ allows us to  derive expression for the derivative of magnetic moment (of conducting electrons) $\vec{M}_e$ with respect to chemical potential
\begin{widetext}
\begin{eqnarray}
	\frac{\partial}{\partial \mu} M_e^i &=&  - \frac{1}{2 V} \epsilon_{ijk} \int d^3 r\, \frac{d^4 p}{(2\pi)^4}\, {\rm tr}\,{\bf G}_W^{(0)} \star \partial_{p_j} {\bf Q}_W^{(0)} \star {\bf G}_W^{(0)} \star \partial_{p_k} {\bf Q}_W^{(0)} \star {\bf G}_W^{(0)} \star \partial_{p_4} {\bf Q}^{(0)}_W(p,r)
\end{eqnarray}
\end{widetext}
Here $V$ is the overall volume of the sample.

We can rewrite this expression as follows
\begin{eqnarray}
	\frac{\partial}{\partial \mu} M_e^i &=& - \frac{1}{4 \pi^2} \, {\cal M}_i
\end{eqnarray}
where ${\cal M}_i$ given by Eq. (\ref{M00}) 
is a topological quantity for the case, when expression standing inside the integral does not have singularities. This is the same expression as the one entering formula for Hall conductivity given in Sect. \ref{QHE}.

In the presence of singularities this quantity is robust to those variations of ${\bf Q}_W$ and ${\bf G}_W$, which vanish in small vicinities of singularities. Suppose that using such a transformation we are able to remove the inhomogeneities completely \footnote{This is a rather strong requirement. However, we suppose that even if it is not fulfilled, the obtained result for the intrinsic magnetization remains valid (then the value of vector $\vec{\beta}$ (defined for the homogeneous system) is to be replaced by  $\langle \vec{\beta}\rangle $, where averaging is over the inhomogeneities)}. Then for the case of the Weyl semimetals with their Weyl points and Fermi surfaces we can reduce the calculation of ${\cal M}$ to that of the homogenous system with ${\bf Q}_W$ and ${\bf G}_W$ depending only on momentum:
\begin{eqnarray}
	{\cal M}_i &=&   \frac{1}{3! \, 4\pi^2\,  V} \epsilon_{ijkl} \int d^3 r\, {d^4 p}\, {\rm tr}\,{\bf G}_W^{(0)}  \partial_{p_j} {\bf Q}_W^{(0)}\nonumber\\ &&  {\bf G}_W^{(0)}  \partial_{p_k} {\bf Q}_W^{(0)}  {\bf G}_W^{(0)} \partial_{p_l} {\bf Q}^{(0)}_W
\end{eqnarray}
 The latter case has been considered in \cite{Z2016_1}. First of all, in case of the non - interacting system with $Q_W = i p_4 - \hat{\cal H}(p)$ we arrive at
 \begin{eqnarray}
 	{\cal M}_i = \frac{\epsilon_{ijk}}{4\pi}\sum_{occupied} \int d^3p {\cal F}_{jk}
 \end{eqnarray}
 where the sum is over the occupied energy levels, while $\cal F$ is Berry curvature. It is instructive to consider the case of rectangular lattice, when the two Weyl points of opposite chirality are placed along the $p_3$ axis  in momentum space at the points $\pm \beta/2$. In this case the third component of ${\cal M}_i$ is nonzero, and is given by
 \begin{eqnarray}
 	{\cal M}_3 &=& \int dp_3 N(p_3), \nonumber\\  N(p_3) &=&  \frac{1}{2\pi}\sum_{occupied} \int dp_1\wedge dp_2\, {\cal F}_{12}(p_1,p_2,p_3)
 \end{eqnarray}
  The value of $N(p_3)$ is integer. We can consider $p_3$ as parameter. Then ${\cal F}_{12}(p_1,p_2,p_3)$ is Berry curvature for the Hamiltonian ${\cal H}(p_1,p_2|p_3)$ depending on $p_3$ as on parameter. For $p_3 \ne \pm \beta$ this is the Hamiltonian of a $2D$ insulator.  $N(p_3)$ is constant $\nu_+$ for $-\beta/2 < p_3 < \beta/2$, and it is equal to another constant $\nu_-$ at $\beta/2 < p_3 $ and $-\beta/2 > p_3$. The non - marginal topologically protected Weyl points correspond to $\nu_+ - \nu_- = -1$ (the left - handed Weyl point is situated at $p_3 = -\beta/2$). Then
  $$
  {\cal M}_3  =  \beta \nu_+ + (2\pi - \beta) \nu_-
  $$
  In the simplest possible case with $\nu_- = 0$ we arrive at
  $$
  {\cal M}_3  =  -\beta
  $$
  The same value is obtained in \cite{Z2016_1} for several tight - binding models. On the basis of these findings we expect that in the general case the value of ${\cal M}$ is given by
  \begin{equation}
  	\vec{\cal M} = -\sum_{k}  \vec{\beta}_k \equiv - \vec{\beta}\label{Mbeta0}
  \end{equation}
where the sum is over the pairs of Weyl points while $\vec{\beta}_k$ is vector in momentum space connecting the Weyl points of the $k$ - th pair.

The above listed results suggest the following expression for the magnetization of Weyl semimetals:
\begin{equation}
	\vec{M}_e = \frac{\mu - \mu_0}{4\pi^2}\sum_{k}  \vec{\beta}_k \label{Mbeta}
\end{equation}
Here $\mu_0$ is the level of chemical potential, at which the magnetization vanishes. For the Weyl semimetal with the same value of chemical potential at the Weyl points the value of $\mu_0$ equals to this value. Correspondingly, $\mu - \mu_0$ is excess of the Fermi energy over the level of Weyl points.

Let us consider the general case of the system with arbitrary Brillouin zone, with the simplification that there are only two Weyl points of opposite chirality. For each point $r$ in coordinate space we consider the foliation of momentum space into the equidistant hyperplanes $\Sigma_\tau$ such that for any value of $\tau$ the given hypersurface crosses the line connecting the given Weyl points. Moreover, we require that the given hypersurface remains inside compact momentum space, which means it is closed through the boundary of the first Brillouin zone. Besides, the two hypersurfaces $\Sigma_\tau$ and $\Sigma_{\tau + \delta}$ are obtained one from another by the shift by $\vec{\bf e} \,\Delta$ in momentum space (here $\vec{\bf e}$ is unity vector in momentum space orthogonal to $\Sigma_\tau$). The requirement of existence of such a foliation restricts the choice of vectors $\vec{\bf e}$. Once such a vector exists, we obtain
\begin{widetext}
\begin{eqnarray}
	\vec{\cal M} \vec{\bf e} &=&   \frac{1}{3! \, 4\pi^2\,  V}  \int d^3 r\, \int d\, \vec{p} \,\vec{\bf e} \, \wedge  \, {\rm tr}\,{\bf G}_W^{(0)} \star d {\bf Q}_W^{(0)} \star \wedge {\bf G}_W^{(0)} \star d {\bf Q}_W^{(0)} \star \wedge {\bf G}_W^{(0)} \star d {\bf Q}^{(0)}_W \nonumber\\&=& \int \, d\, \tau \, {\cal N}_{\bf e}(\tau)
\end{eqnarray}
with
\begin{eqnarray}
	{\cal N}_{\bf e}(\tau) &=&   \frac{1}{3! \, 4\pi^2\,  }  \,\int_{\Sigma_\tau} \, \frac{1}{V}\int d^3 r  \,  {\rm tr}\,{\bf G}_W^{(0)} \star d {\bf Q}_W^{(0)} \star \wedge {\bf G}_W^{(0)} \star d {\bf Q}_W^{(0)} \star \wedge {\bf G}_W^{(0)} \star d {\bf Q}^{(0)}_W
\end{eqnarray}
\end{widetext}
This is the topological invariant for any fixed value of $\tau$, except for the values, where expression standing inside the integral over $\Sigma_\tau$ contains singularities. The latter values of $\tau$ correspond to Weyl points. Let the values of $\tau$ vary between $0$ and $P_{\bf e}$ for certain $P_{\bf e}$.

Recall that for any choice of the Brillouin zone there exists only discrete number of appropriate vectors $\vec{\bf e}$. Say, for the case of the rectangular Brillouin zone there are three such directions, while for the Brillouin zone, which is given by the product of hexagon and the finite interval there are $7$ admitted directions of $\bf e$.

Let $A$ and $B$ be the positions of the Weyl points, then we denote by $\vec{AB}$ vector (of a smaller length) connecting these two points, while $\vec{BA}$ is vector (of a larger length) that connects them. Then for the given admitted vector $\vec{\bf e}$ we obtain
\begin{eqnarray}
	{\cal N}_{\bf e}(\tau) &=& N_{{\bf e},+}, \quad \tau  \vec{\bf e}\in  \vec{AB} \nonumber\\
	{\cal N}_{\bf e}(\tau) &=& N_{{\bf e},-}, \quad \tau  \vec{\bf e}\in  \vec{BA}
\end{eqnarray}
while
\begin{equation}
	\vec{\cal M} \vec{\bf e} = \frac{\mu - \mu_0}{4\pi^2}(N_{{\bf e},+} \vec{AB} + N_{{\bf e},-} \vec{BA}) \vec{\bf e}
\end{equation}
Let us consider the non - marginal Weyl point corresponding to the minimal values of the topological invariant
	\begin{widetext}
	\begin{eqnarray}
		\mathcal{N}_3
		&=&\frac{1}{24 \pi^2 {V}}
		\int d^3x \int_{\Sigma(x)}
		\tr \Bigg[
		{\bf G}_W^{(0)}\star d {\bf Q}_W^{(0)} \star {\bf G}_W^{(0)}
		\wedge \star d {\bf Q}_W^{(0)}\star {\bf G}_W^{(0)} \star \wedge d {\bf Q}_W^{(0)}
		\Bigg]\label{N3}
	\end{eqnarray}
\end{widetext}
Here integration is over the surface surrounding the {\it given} Weyl point (depending on $x$). This topological invariant protects Weyl points. For its minimal values $\pm 1$ we have
$$
N_{{\bf e},+} - N_{{\bf e},-} = -1
$$
for any admissible $\vec{\bf e}$.

Let us consider the simplest situation with vanishing value of $N_{{\bf e},-}$. (In this case ${\cal N}_{\bf e}(\tau)$ vanishes being calculates along the longest path connecting the two Weyl point, while it is equal to unity being calculated along the shortest path.) Then
\begin{equation}
	\vec{\cal M}  = -\vec{AB}
\end{equation}
Recall that this vector connects Weyl points of opposite chirality. It is directed from the left - handed Weyl point to the right - handed one. 

This result can easily be generalized to the case when several pairs of Weyl points are present in the Brillouin zone by Eq. (\ref{Mbeta}), where
the sum is over the pairs of Weyl points while $\vec{\beta}_k$ is the shortest vector connecting the $k$ - th pair. {\it Our convention is that vector $\vec{\beta}_k$ points from the left - handed to the  right - handed Weyl point. } 

It is instructive to compare this "free" electrons contribution to magnetic moment to its experimentally measured total value. For example, in $Co_3 Sn_2 S_2$ there are three pairs of Weyl points. The value of $|\sum_{k}  \vec{\beta}_k|$ depends on total magnetization and is of the order of  $0.85 \times 10^{10}$ $\hbar/m$ at its maximal value (extracted from Fig. 6 of Supplementary material to \cite{wang2018large}).
Density of $Co_3Sn_2S_2$ crystal is $7.24$ g/cm$^3$. Mass of the f.u. is $479\times 1,66\times 10^{-27}$ Kg. The experimental value of saturating magnetic moment is $0.29 \mu_B/Co = 0.29\times 3 \mu_B/$f.u. In SI units we obtain
$$
|\vec{M}| =  0.073 \times 10^6 \, A/m,
$$
which corresponds to the magnetic fields of the order of $0.1$ Tesla.
This value is to be compared to that of Eq. (\ref{Mth}). In the system of units SI Eq. (\ref{Mbeta}) receives the form:
\begin{equation}
	\vec{M}_e = \frac{e}{h}\frac{\mu - \mu_0}{2 \pi \hbar}\sum_{k}  \vec{\beta}_k \label{Mth2}
\end{equation}
Excess of Fermi energy over the Weyl point energy is $60$ meV $= 60\times 10^{-3}\, 1.6\, 10^{-19} $ J, and we arrive at
$$
|\vec{M}_e| =  3.14\times 10^3\, A/m \approx 0.01 \, \mu_B/Co
$$
This value is about $25$ times smaller than the total saturating magnetic moment.

\subsection{Persistent electric current on the surface of the sample and intrinsic anomalous Hall effect calculated along the boundary}
	
Along the boundary of the two - dimensional insulator the persistent electric current flows. In the presence of difference in chemical potentials at the opposite sides of the sample the currents along these sides do not cancel each other, which results in the Hall current. In the presence of sufficient amount of disorder the Hall current flows only along boundary.

In the magnetic Weyl semimetal the situation is similar. Let us consider for simplicity the sample of rectangular form with one of its sides (of length $L$) directed along vector $\vec{\beta} \equiv \sum_k \vec{\beta}_k$, the other sides are of lengths $H$ and  $R$ (see Fig. \ref{fig0}). At the surface of the sample there are the Fermi arcs connecting Weyl points. For simplicity we can assume that the only Fermi arc is along the straight line, and the surface Fermi velocity is constant along the Fermi arc. (The result for the general case will be the same.) In this situation the contribution of Fermi arcs to electric current is calculated as follows:
\begin{equation}
	J_{arc} = \frac{|\vec{\beta}|\, \delta p}{(2\pi)^2} v^{(s)}_F\, L = \frac{|\vec{\beta}|\, \mu}{(2\pi)^2} L = |\vec{M}_e| L
\end{equation}
Here $\delta p$ is thickness of the area in momentum space close to the Fermi arc occupied by the quasiparticles. $v^{(s)}_F$ is surface Fermi velocity. The product $\delta p\, v^{(s)}_F$ is equal to chemical potential $\mu$. One can see that the total magnetization is given by
$J RH/V$ as it should (here $V = RHL$).

In case of the saturating magnetic moment $M \approx 0.29 \mu_B/Co$ and the sample of length $\approx 0.5$ cm we obtain the contribution to persistent current of the Fermi arcs:
$$
J_{arc} \approx  16 \, A
$$

Let us now consider the case, when between the two opposite sides (parallel to the direction of $\vec{M}_e$) there is a voltage $U$. In this case the intrinsic (Fermi arc) contribution to the total Hall current that flows along the boundary of the sample is given by
\begin{equation}
	J^{intrinsic}_H = \frac{|\vec{\beta}|\, U}{(2\pi)^2} L = \frac{|\vec{\cal M}| L}{(2\pi)^2}
\end{equation}
We come to the following value of intrinsic Hall conductivity:
$$
\sigma^{intrinsic}_H = \frac{|\vec{\cal M}|}{2\pi \hbar} \, \frac{e^2}{h}
$$
(Here we restore the SI units.)
We will see in Sect.  \ref{AQHE} that the same expression is obtained using the bulk description. Notice that in addition to this contribution to Anomalous Hall conductivity there is the contribution due to the scattering processes (see Sect. \ref{AQHE}).

\subsection{Contribution of intrinsic AQHE to the total anomalous Hall resistivity}
	\label{AQHE}
	There are three sources of Hall current: external magnetic field $\vec{H}$, magnetization $\vec{M}$, topological contribution originated from Weyl points and related to the difference in positions of Weyl points $\vec{\beta} \equiv\sum_k \vec{\beta}_k$ .
	 Relation between these quantities is
	\begin{equation}
		\vec{E} = \Big(R_0 \mu_0 \vec{H}  + R_1 (\vec{M} -\vec{M}_e)  + R^\prime_1 \vec{M}_e\Big) \times \vec{j} + \rho_{\parallel} \vec{j}  \label{ohm}
	\end{equation}
where $\rho_\parallel$ is ordinary longitudinal resistivity, $R_0$ is ordinary Hall constant, $R_1$ is the extraordinary Hall constant (in contrast to the ordinary one). $R_1^\prime$ is the intrinsic AQHE constant.

Let us consider for the moment the idealized situation, when external magnetic field is absent $\vec{H}=0$, the whole magnetization is equal to the intrinsic one $\vec{M} = \vec{M}_e\ne 0$, while $\rho_\parallel \to 0$ (scattering of charge carriers is neglected)\footnote{According to the data of \cite{wang2018large} this situation is not realistic, at least, for Co$_3$Sn$_3$S$_2$, where the intrinsic magnetization does not vanish only for the sufficiently large value of total magnetic moment, which means that always $|\vec{M}| \gg |\vec{M}_e|$. Nevertheless we consider this situation in order to separate the topological contribution to AQHE.}. Using results  of \cite{Z2016_1} we would then obtain the following relation (see Sect. \ref{QHE}):
 \begin{equation}
 	\vec{j} = -\frac{1}{4\pi^2}  \vec{\cal M} \times \vec{E}\label{jME}
 \end{equation}
while according to Sect.  \ref{Magnetic} $\vec{\cal M}$ is given by Eq. (\ref{Mbeta0}). 
For the intrinsic anomalous conductivity we would obtain
\begin{equation}
	\sigma^{ik}_{H} = \frac{\epsilon^{ijk}}{2\pi\hbar} \Big(\sum_{q}  \vec{\beta}^{j}_q\Big)\, \frac{e^2}{h} \label{sigmapr}
\end{equation}
where we restored the electron charge and Plank constant.

On the other hand,
\begin{equation}
	\vec{E} = {R}^\prime_1 \vec{M}_e \times \vec{j} + \gamma \vec{M}_e
\end{equation}
Here $\gamma$ accounts for the appearance of electric field component along the magnetic moment induced by the band topology.
Therefore, 	
\begin{eqnarray}
	\vec{E} &=& R^\prime_1 \frac{\mu - \mu_0}{4\pi^2}  \vec{\beta} \times \Big[\frac{1}{4\pi^2}    \vec{\beta} \times \vec{E}\Big] + \gamma \frac{\mu - \mu_0}{4\pi^2}  \vec{\beta} \nonumber\\ &=&   R^\prime_1 \frac{\mu - \mu_0}{(2\pi)^4}  \vec{\beta} (\vec{E}     \vec{\beta})   -  R^\prime_1 \frac{\mu - \mu_0}{(2\pi)^4}  \vec{\beta}^2   \vec{E} \nonumber\\&&+ \gamma \frac{\mu - \mu_0}{4\pi^2}  \vec{\beta}
\end{eqnarray}
We obtain the following expression for the Hall constant:
\begin{equation}
	R^\prime_1 = -\frac{(2\pi)^4}{(\mu - \mu_0)\Big(\sum_{k}  \vec{\beta}_k\Big)^2 }
\end{equation}
The value of $\gamma$ cannot be predicted basing on Eq. (\ref{jME}) and remains arbitrary. We choose it equal to zero.

In practise Eq. (\ref{sigmapr}) cannot be applied because the ordinary magnetization as well as ordinary resistance do not vanish. Let us consider now the more general case, when $H=0$, while $M\ne 0$, $\rho_\parallel\ne 0$. Then we should return to Eq. (\ref{ohm}), which may be rewritten as
\begin{equation}
	\vec{E} = \Big( R_1 (\vec{M}-\vec{{ M}_e}) + {R}^\prime_1 \vec{ M}_e\Big) \times \vec{j} + \rho_{\parallel} \vec{j}  \label{ohm2}
\end{equation}
In this case Eq. (\ref{ohm2}) is reduced to
\begin{equation}
	\vec{E} =  \rho_\bot \hat{\vec{k}}  \times \vec{j} + \rho_{\parallel} \vec{j} = \hat{\rho} \vec{j} \label{ohm3}
\end{equation}
where $\hat{\vec{k}}$ is a unity vector while
$$
\rho_\bot \hat{\vec{k}} =   R_1 \vec{M}+  ({R}^\prime_1-R_1) \vec{M}_e
$$
Here
\begin{eqnarray}
	{\rho}^{ik} = \rho_\bot\epsilon^{ijk}\hat{k}_j + \rho_\parallel \delta^{ik}
\end{eqnarray}
Let axis $y$ be directed along $\vec{k}$. Then
$$
\hat{\rho} = \left(\begin{array}{ccc}\rho_\parallel & 0 & \rho_\bot \\
	0 &  \rho_\parallel & 0\\
-\rho_\bot & 0 & \rho_\parallel  \end{array} \right)
$$
We represent electric current as
$$
\vec{j} = \hat{\sigma}\, \vec{E}
$$
where $\hat{\sigma}$ is the conductivity matrix given by
\begin{eqnarray}
\hat{\sigma} = \left(\begin{array}{ccc}\frac{\rho_\parallel}{\rho_\bot^2 + \rho_\parallel^2} &  0 &  \frac{\rho_\bot}{\rho_\bot^2 + \rho_\parallel^2}  \\
	0 &  1/\rho_\parallel & 0 \\
-	\frac{\rho_\bot}{\rho_\bot^2 + \rho_\parallel^2} & 0 & \rho_\parallel  \end{array} \right)	
\end{eqnarray}
For the Hall conductivity we obtain
\begin{eqnarray}
	\sigma_H &=& \frac{|R_1 (\vec{M}-\vec{M}_e)-  \frac{(2\pi\hbar)^2\sum_{q}  \vec{\beta}_q}{e^2(\sum_{k}  \vec{\beta}_k)^2}|}{(R_1 (\vec{M}-\vec{M}_e)-  \frac{(2\pi\hbar)^2\sum_{q}  \vec{\beta}_q}{e^2(\sum_{k}  \vec{\beta}_k)^2})^2 + \rho_\parallel^2}\label{sigmaR}	
\end{eqnarray}
(Here we restore SI units.)

This expression might be compared to experiment. For example, in $Co_3 Sn_2 S_2$ there are three pairs of Weyl points. The value of $|\vec{\beta}| = |\sum_{k}  \vec{\beta}_k|$ is about $0.85 \times 10^{10}$ $m^{-1}$ at the saturating value of total magnetization and decreases with decrease of the total magnetization (see, for example \cite{wang2018large}). We obtain for the would be value of Hall conductivity (calculated with $M$ and $\rho_\parallel$ neglected)
$$
\sigma^{intrinsic}_{xz} \approx 0.52\,\times 10^{3} \Omega^{-1}\, cm^{-1}
$$
which is about twice smaller than the experimental value of Hall conductivity announced in \cite{liu2018giant} for temperature $3K$ and $M \approx 0.29 \mu_B/Co$. For this value of magnetization and temperature around $3$K the value of ordinary conductivity is $\rho_\parallel \approx 75 \mu \Omega\, cm$. Besides, it is known that $R_1$ is proportional to $\rho_\parallel^2$:
$$
R_1 = \lambda \rho_\parallel^2
$$
which means it is dominated by the side jumps contribution.
The value of $\lambda$ may be extracted from Eq. (\ref{sigmaR}).

We can see that although the contribution of band topology to magnetization is around $4$ per cent, its contribution to the Hall conductivity is around $40$ per cent, which means that it is the dominated contribution to anomalous Hall effect.

\section{Weyl orbits as probe of Chiral separation effect}

\label{WeylOrbit}

\begin{figure}[h]
	\centering  %
	\includegraphics[width=0.9\linewidth]{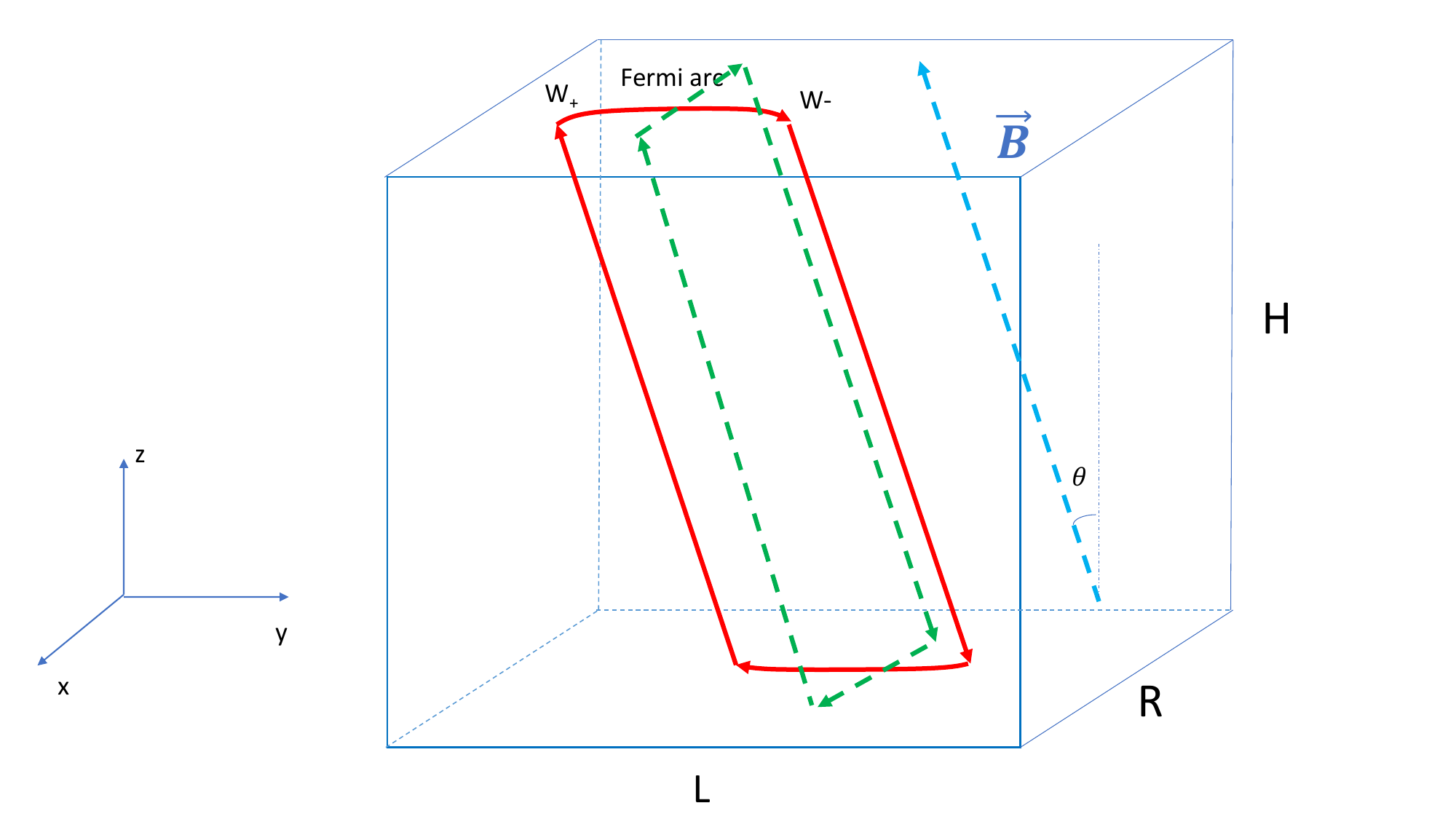}  %
	\caption{The sample of Weyl semimetal of hight $H$, width $R$ and length $L\gg H$. On the upper side of the sample the Fermi arc is drawn schematically. The position of the right - handed Weyl point W$_+$ is connected by the Fermi arc with the position of the left - handed Weyl point W$_-$. Weyl points of the same chirality situated on the opposite sides of the sample are connected by the solid lines that represent schematically motion of quasiparticles in the bulk. Dashed closed line  represents motion of quasiparticles in coordinate space along Weyl orbit.   }  %
	\label{fig2}   %
\end{figure}

\subsection{Chiral separation effect and the flow of states along the Fermi arcs}
	
	The chiral separation effect results in axial current along external magnetic field (Sect. \ref{CSE}, see also  \cite{zubkov2023effect}). The corresponding conductivity is given by the topological invariant that is robust to introduction of interactions and disorder.   Let us consider the rectangular sample of magnetic Weyl semimetal with one of its sides (of length $L$) directed along the axis of  $\vec{\beta} = \sum_k\vec{\beta}_k$ (this axis is determined by the crystal structure). In case of nonzero magnetization of the sample its component  $\vec{M}_e$ (caused by band topology) is directed along $\vec{\beta}$. In addition, we impose external magnetic field in plane $(yz)$. It causes inside the sample magnetic induction field $\vec{B}$, which is directed along vector of magnetization $\vec{M}$. Angle between its direction and axis $z$ is $\theta$, see Fig. \ref{fig2}.  {\it Magnetic induction $\vec{B}$ is actually the microscopic magnetic field averaged over physically small regions. It should be distinguished from magnetic field strength $\vec{H}$ in substance. However, in the following we will refer sometimes to field $\vec{B}$ simply as to magnetic field if this will not cause ambiguity.}  $|\vec{\beta}|$ depends on the total magnetization and on angle $\theta$. It reaches maximum at $\theta = \pi/2$ and for the saturating value of magnetization. Dependence of $|\vec{\beta}|$ on $\theta$ is not well - known. However, it is more or less obvious that in order to reach meaninfull values of $|\vec{\beta}|$ we should keep $\theta$ not too small. At the same time we do require that this angle differs from $\pi/2$ in order to observe effects caused by the $z$ component of magnetic induction $\vec{B}$.  Besides, we assume that the value of magnetic induction does not differ much from the value
	$$
	B \sim \mu_0 \times |\vec{M}| \approx  0.1 \, T,
	$$
	which corresponds to maximal possible value of magnetization. 
	
	The hight of the sample is $H$, while the width is $R$. Length $L$ of the sample is assumed to be much larger than $H$ in order to avoid consideration of side effects.  
	
	Magnetic induction $\vec{B}$ results in the bulk axial current. It flows between the opposite sides of the sample. Since Weyl points of opposite chirality are separated in momentum space, the axial current actually means the flow along magnetic field of the quasiparticles from the vicinity of the left - handed Weyl point, and the counterflow in oppisite direction of the quasiparticles from the vicinity of the right - handed Weyl point. The axial current that crosses horizontal section of the sample is given by
	\begin{equation}
		{J}_5 = \frac{\cal N}{(2\pi)^2}2\mu {B} LR \,{\rm cos}\, \theta
	\end{equation}
	Here $\cal N$ is the number of pairs of Weyl fermions (equal to $3$ for the case of Co$_3$Sn$_2$S$_2$). In the absence of external electric field the axial current is conserved, which means that it transfers the chiral charge (equal to the difference between the quasiparticles with opposite chirality).
	
	The above expression for the axial current can be obtained also from the following representation of axial current density:
	$$
	j_5 = 2 {\cal N}\, n(\mu)\, v_F,
	$$
	where $n(\mu) = \frac{B\mu}{v_F 4\pi^2}$ is the density of electrons in the presence of magnetic induction field $B$ (that reside close to one Weyl point with chemical potential $\mu$).
	
	\mzz{As a result, in the bulk the number of the right - handed quasiparticles that cross the section of the sample during the unit of time is given by $\dot{N}_+ = \frac{\cal N}{(2\pi)^2}\mu |\vec{B}| LR\,{\rm cos}\, \theta$, while the number of the left - handed quasiparticles that cross the section during the unit of time is  $\dot{N}_- = - \frac{\cal N}{(2\pi)^2}\mu |\vec{B}| LR\,{\rm cos}\, \theta$. This process is accompanied  by the flow of the states along the Fermi arcs \cite{Potter_2014} with the same rate.} This flow occurs due to the Lorentz force. For simplicity let us consider again the case, when the only Fermi arc is along the straight line. Equation that governs the flow along the Fermi arc is
	$$
	\dot{p} = v_F B\,{\rm cos}\, \theta
	$$
	The number of particles that cross during the unit of time the given section of the Fermi arc is
	$$
	\dot{N} = \frac{v_F\, B \,\delta p}{(2\pi)^2} LR\,{\rm cos}\, \theta = \frac{\mu \, B}{(2\pi)^2} LR  \,{\rm cos}\, \theta = \dot{N}_+ = -\dot{N}_-
	$$
	\mzz{One can see that in this particular case the rate of the flow along the Fermi arc equals to the rate of the chiral separation effect in the bulk.} This conclusion may obviously be extended also to the more complicated configuration of the Fermi arcs and Weyl points. It has been proposed in \cite{Potter_2014} that the closed path in phase space is quantized (that contains sliding along the Fermi arcs of opposite sides and moving between these sides of particles that reside near the Weyl points of opposite chiralities). The hypothesis has been proposed that "motion" along the given closed orbit is quantized, and the semi - classical quantization procedure has been applied.

\subsection{Semi - classical quantization of motion along Weyl orbit}

Weyl orbit is represented schematically on Fig. \ref{fig2}. In coordinate space the semi - classical trajectory has the rectangular form  represented by the dashed line. Its bulk part is described in Appendix \ref{clas}. In quantum theory it corresponds to the chiral lowest Landau level (LLL). Horizontal part is caused by motion of quasiparticles along the upper (lower) side of the sample (which in momentum space reside close to the Fermi arc).

	For the sample of the geometry described above the semi - classical quantization reads
\begin{equation}
	\oint \vec{p}  d\vec{r}  = 2\pi (n + \gamma),
\end{equation}
Let us denote $\pi = p - e A(r)$. In terms of the new variables ($\pi, r$) we obtain
\begin{equation}
	\oint \Bigl(\vec{\pi} + e \vec{A}(r)\Bigr) d\vec{r}  = 2\pi (n + \gamma),
\end{equation}
where  $n$ is an integer number while $\gamma$ is unknown real number depending on the details of the system.

For the classical trajectory in the bulk angular momentum (projected to the direction of magnetic field) is given by
$$
L = r \, \pi_\bot = \frac{\pi_\bot^2}{eB} = 2\pi m
$$
where $m$ is integer. Correspondingly, the values of $\pi_\bot$ and $r$ are quantized:
$$
\pi_\bot = \sqrt{2\pi m \,e B}, \quad r = \sqrt{\frac{2\pi m}{e B}}
$$
The Lowest Landau Level (LLL) corresponds to $m = 0$. In this situation velocity is directed along magnetic field and is equal to $v_F$.

One of the two bulk parts  of the orbit represented in Fig. \ref{fig2} contributes the adiabatic invariant entering the semi - classical quantization condition by (see Appendix \ref{clas}):
\begin{eqnarray}
	I_h &=&   \frac{\epsilon H}{v_F\, {\rm cos}\, \theta}
\end{eqnarray}
where $\epsilon = v_F |\vec{\pi}|$ is energy of the particle.
Notice, that there is no rotational part of motion for the states that belong to the LLL (which is the case of interest for us).

The state with the given value of momentum along magnetic field is degenerate: the degeneracy is given by
$\frac{eB}{4\pi^2\hbar^2 } A $, where $A$ is area of the sample in direction orthogonal to magnetic field.

When quasiparticle reaches the upper side of the sample, it encounters the Hamiltonian of the Fermi arc, which equals to zero along the Fermi arc (let it be directed along the $y$ axis). The derivative of the Hamiltonian along the $x$ axis is equal to surface Fermi velocity $v_F^{(s)}$. We can simply consider $H(p,r) = - v^{(s)}_F \pi_x$. (We omit for simplicity the consideration of the end points of the Fermi arc.) Action has the form
\begin{eqnarray}
	S &=&
	\int \Big( \pi_x d x + \pi_y dy - e B x d y \, {\rm cos}\, \theta - v^{(s)}_F \pi_x dt\Big)
\end{eqnarray}
Classical equations of motion read
\begin{eqnarray}
	&&d x = v^{(s)}_F dt\nonumber\\ &&
	d y = 0\nonumber\\ &&
	0 = - d\pi_x - eBdy \, {\rm cos}\, \theta\nonumber\\ &&
	0 = -d\pi_y + eB dx\, {\rm cos}\, \theta
\end{eqnarray}
Solution of these equations is:
\begin{eqnarray}
	y &=& const\nonumber\\
	\pi_x & = & p_x = const\nonumber\\
	\pi_y & = & const + e B v_F^{(s)} t\, {\rm cos}\, \theta\nonumber\\
	x & = & const + v_F^{(s)} t
\end{eqnarray}
The value of $\pi_x$ for the given trajectory is given by $\epsilon/v_F^{(s)}$ and belongs to the interval between $0$ and $\mu/v_F^{(s)}$. The value of $\pi_y$ belongs to interval between $0$ (the position of W$_+$) and $\beta$ (the position of W$_-$). The contribution to the adiabatic invariant corresponding to the given Fermi arc is
\begin{eqnarray}
	I_L &=& \int \Big(\pi_x dx + \pi_y dy  + e A_x(x,y) dx + e A_y(x,y) d y  \Big)\nonumber\\
	& = & \frac{\pi_x v_F^{(s)}\beta}{e B v_F^{(s)}\, {\rm cos}\, \theta}  = \frac{\epsilon \beta}{e B v_F^{(s)}\, {\rm cos}\, \theta}
\end{eqnarray}
The total quantization condition is
\begin{equation}
2\frac{\epsilon \beta}{e B v_F^{(s)}\, {\rm cos}\, \theta} + 2 \frac{\epsilon H}{v_F\, {\rm cos}\, \theta}	= 2\pi \hbar (n + \gamma)
\end{equation}
This results in the following expression for the quantized energy levels:
\begin{equation}
	\epsilon_n = \frac{\pi \hbar v_F(n+\gamma)\, {\rm cos}\, \theta}{H + \frac{\beta}{e B} \frac{v_F}{v_F^{(s)}}}, \quad n = 0, 1, 2,...
\end{equation}
This result coincides with that of \cite{zhang2016quantum}. For $\theta = 0$ this expression is reduced to the result obtained in \cite{Potter_2014}. This consideration indicates that the corresponding values of energies might be observed as frequencies of quantum oscillations in various observed quantities. In \cite{Potter_2014} it was proposed to observe these oscillations in experiment. Observation of these oscillations would represent the indirect detection of chiral separation effect.  Below, in Sect. \ref{prop} we propose different ways to detect experimentally the bulk CSE, which are also inderect but seem to us more straightforward and easy for implementation.

\subsection{Magnetic moment resulted from the Weyl orbit }
	
Let us calculate contribution of Weyl orbits to magnetic moment. For this reason we calculate the electric current along the upper side of the sample (along axis $x$). This current, in turn, is equal to the number of right - handed quasiparticles per unit time that are incoming from the bulk.  The latter number is equal to the half of the axial current, and it may be calculated as the product of the number of quasiparticles (participated in the current) and their Fermi velocity. We should take into account that after the current passes distance $l$, it turns down. Overall, in order to calculate the electric current we should multiply the current coming from the bulk by the ratio $l/R$:
\begin{equation}
	J =\frac{l}{R} \frac{eB}{4\pi^2\hbar^2 }\,LR\, \frac{\mu}{v_F}\,e\,v_F\,{\rm cos}\,\theta  = \frac{e^2 B \mu\, l}{4\pi^2 \hbar^2 }\,L\,{\rm cos}\,\theta
\end{equation}	
The same electric current flows along the upper side of the sample in the $x$ direction. However, it passes the limited length equal to $l = \frac{v_F^{(s)} \beta}{e B v_F^{(s)}\,{\rm cos}\,\theta} = \frac{\beta}{e B \,{\rm cos}\,\theta}$ (see Fig. \ref{fig2}) with $\beta = |\vec{\beta}|$. Correspondingly, the induced magnetic moment in $y$ direction (per unit volume) is equal to
\begin{eqnarray}
	J R H &=&\frac{e^2 B (\mu-\mu_0)}{4\pi^2 \hbar^2 }\,l \,{\rm cos}\,\theta\,  = \frac{e (\mu-\mu_0) \beta}{4\pi^2 \hbar^2 }\, \nonumber\\& =& \frac{e (\mu-\mu_0) \beta}{4\pi^2 \hbar^2 }\,  = M_e
\end{eqnarray}
We restore here the  energy level of Weyl points $\mu_0$. The final expression is equivalent to that of the band contribution to magnetic moment (calculated above in Sect. \ref{Magnetic}) $M_e$ as expected, which means that the motion of electron along the Fermi arcs forms electric current related to the internal magnetic moment. We performed this calculation as if there was only one pair of the Weyl points. Obviousely, in the presence of several pairs the final answer is not changed.

Let us estimate the size of the Weyl orbit in $x$ direction  (in the direction orthogonal to  the magnetic axis $\vec{\beta} = \sum \vec{\beta}_k$ and to the direction of external magnetic field) for the case of the sample of Cd$_3$Sn$_2$S$_2$ with  $B = 10^{-1} $ T. Then
$$
l  \approx 6\times 10^{-5}\, m/\,{\rm cos}\,\theta = 0.06\, mm/\,{\rm cos}\,\theta
$$
This value is to be compared with the magnetic length $l_B = \sqrt{\hbar/eB} \sim  10^{-7} $m and the sizes of the sample.

\section{Proposal for the experimental detection of Weyl orbits in the presence of external electric field}
\label{prop}

\begin{figure}[h]
	\centering  %
	\includegraphics[width=0.9\linewidth]{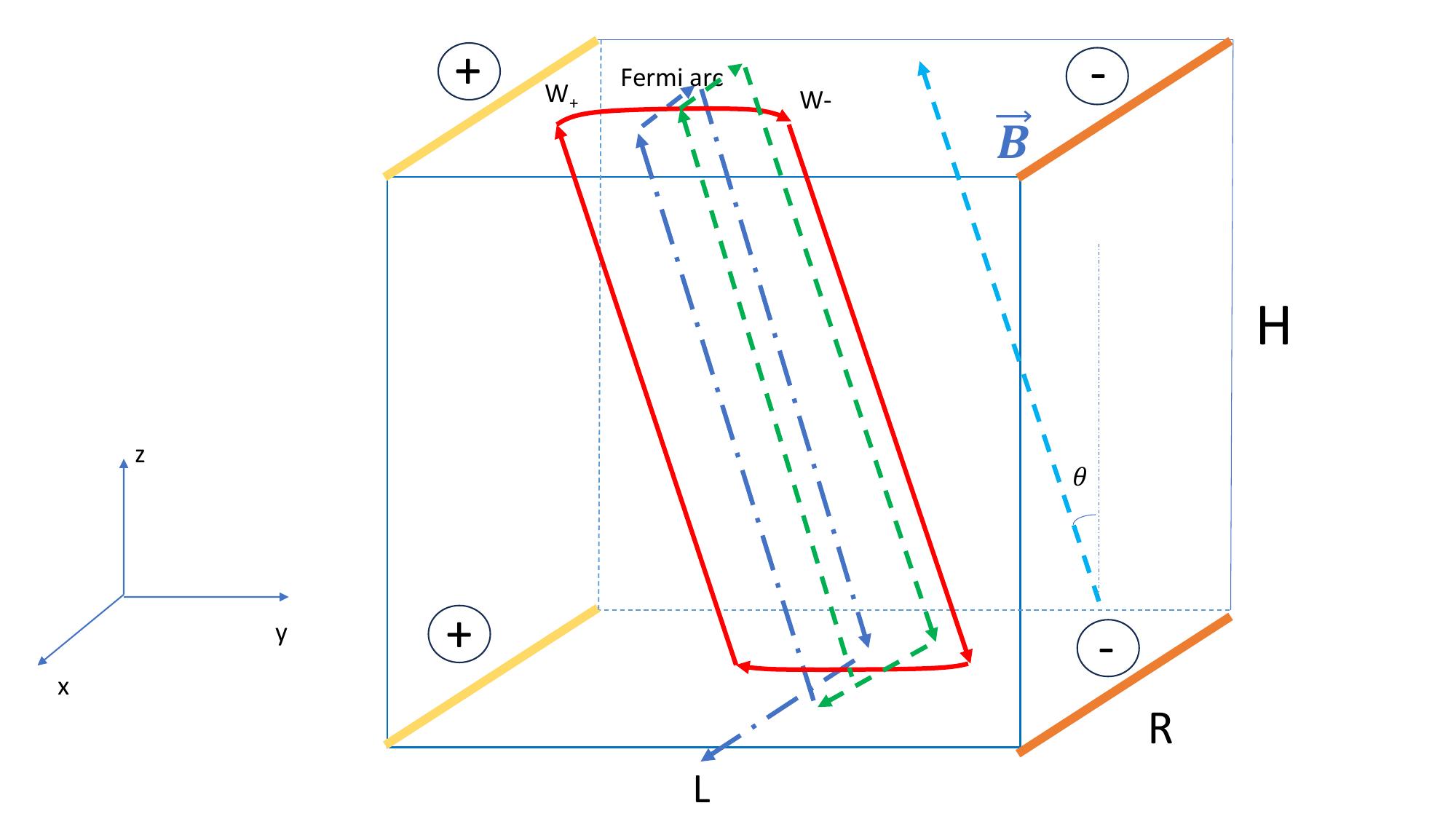}  %
	\caption{Cycling dashed lines  represent motion of quasiparticles in coordinate space along Weyl orbits. In addition, it is assumed that electric potential is distributed in a non - trivial way along the surface of the sample. The resulting electric field is directed from $"-"$ to $"+"$ between left and right sides of the sample. As a result the Weyl orbit is not closed. Its length on the upper side becomes smaller than the length on  the lower side. The first cycle of the Weyl orbit is represented by the dashed line while the second cycle is represented by the dashed - dotted line. }  %
	\label{fig5}   %
\end{figure}	


	Let us consider the situation when  the homogeneous electric field $E$ is added between the two opposite sides of the sample. In order to apply electric field  we may add the leads, which maintain constant difference in potentials as shown schematically in Fig. \ref{fig5}. This difference in potentials, first of all, will give rise to the ordinary electric current. As well as in the other experiments with Weyl semimetals in order to observe the topological properties, one should keep sufficiently low temperatures (of a few K).

 In addition, voltage between the left and the right sides of the sample results in the Hall effect: the electric current directed along axis $x$. (Or, inverse: the electric current along axis $x$ results in Hall voltage between the sides.) Below we will discuss the topological contribution to the Hall current caused by the Weyl orbits. 
 
 Besides, the electric field directed along the internal magnetic moment (along axis $y$) gives rise to the tunneling in the bulk between the two Weyl points, i.e. the appearance of the chiral imbalance. This imbalance, in turn, together with the magnetic field directed along the $z$ axis through the (nonequilibrium) chiral magnetic effect results in the electric current along $z$ axis. This latter effect is assumed to be subdominant, we do not discuss it here. Instead we concentrate on the effect of electric field on the upper and lower sides of the samples, where it modifies sliding along the Fermi arcs.
 
 In the presence of electric field the states corresponding to Landau levels drift along axis $x$. Consideration of this drift gives the simplest way to evaluate Hall conductance. It does not provide the exhaustive calculation since scattering of the quasiparticles is to be taken into account. However, this method provides the order of magnitude estimate.  Later we will compare the result obtained this way to the contribution of Weyl orbits in order to find regime, in which the latter contribution becomes valuable. We consider Weyl fermions in the presence of vertical magnetic induction field $B_\bot$ (the value of $B_\parallel$ parallel to $\vec{\beta}$ is disregarded in this estimate). At zero temperature, and with the interactions disregarded all quasiparticles with energy above the level of Fermi points participate in the Hall current. We perform the (emergent) Lorentz transformation to the reference frame moving with velocity $E/B_\bot$ (we consider the case of suffuciently weak electric field, when this ratio is much smaller than $1$). In this reference frame effectively there is no elecric field acting on the quasiparticles, and the states of LLL are occupied up to $p_z = (\mu - \mu_0)/v_F$. This gives electric current  density 
 $$
 j = 2\,e\, {\cal N}\, n(\mu)\, E/B_\bot,
 $$
 where $n(\mu) =e \frac{B_\bot(\mu - \mu_0)}{v_F 4\pi^2 \hbar^2}$ is the density of electrons at each Weyl point. From here we derive expression for the conductivity
 $$
 \sigma^{(0)}_{xy} =2 e^2{\cal N} \frac{B_\bot(\mu-\mu_0)}{v_F 4\pi^2\hbar^2} \frac{1}{B_\bot} = \frac{2 e^2 {\cal N} (\mu-\mu_0)}{v_F 4\pi^2 \hbar^2}
 $$ 
 Multiplying by the area of the sample section we obtain the following expression for conductance:
 $$
 \Sigma^{(0)}_{xy} =\frac{2 {\cal N} (\mu-\mu_0)}{ v_F h}\frac{e^2}{h} HL
 $$
 
Now let us consider the specific contribution of Weyl orbits to Hall current. The length of the Weyl orbit in $x$ direction is changed because the force acting on the quasiparticle receives contribution from electric field. This length is decreased on the upper side while it is increased on the lower side.
	
	On the upper side we obtain
	$$l_+ = \frac{v_F^{(s)} \beta}{e B_\bot v_F^{(s)} + e E} = \frac{\beta}{e B_\bot }\frac{1}{1 + \frac{E/B_\bot}{v_F^{(s)}}}$$
	while on the lower side
		$$l_- = \frac{v_F^{(s)} \beta}{e B_\bot v_F^{(s)} - e E} = \frac{\beta}{e B_\bot }\frac{1}{1 - \frac{E/B_\bot}{v_F^{(s)}}}$$
	(It is worth mentioning that we consider here the case of small electric field with $E/B_\bot < v_F^{(s)}$.) 
	One can see that such a modification of the form of the Weyl orbit results in electric current along axis $x$. Its value may be evaluated as follows. The overall time needed for the particle to move along one cycle of the Weyl orbit is given by
	$$
	\tau = \frac{l_+}{v_F^{(s)}} + \frac{l_-}{v_F^{(s)}} + 2\frac{H}{v_F}
	$$
	The density of bulk quasiparticles that participate in the Weyl orbits (of one pair of Weyl points) is
	$2 n =  2 \frac{eB_\bot}{4\pi^2\hbar^2 }\, \frac{(\mu - \mu_0)}{v_F} $. As a result of sliding along the modified Weyl orbit each particle moves in average in the direction of $x$ axis with velocity
	$$
	V = \frac{l_- - l_+}{\tau}
	$$
	and we arrive at the following expression for the contribution to electric current
	\begin{equation}
		J^{(bulk)}_x = 2 \frac{ e^2B_\bot(\mu - \mu_0)}{4\pi^2\hbar^2 } \frac{\frac{v_F^{(s)}}{v_F}\Big( \frac{1}{1 - \frac{E/B_\bot}{v_F^{(s)}}} - \frac{1}{1 + \frac{E/B_\bot}{v_F^{(s)}}}\Big)LH }{\frac{1}{1 - \frac{E/B_\bot}{v_F^{(s)}}} + \frac{1}{1 + \frac{E/B_\bot}{v_F^{(s)}}} + 2\frac{v_F^{(s)}e H B_\bot}{v_F \beta}  }
	\end{equation}
 	Let us consider the regime, in which $E/B_\bot$ is much smaller than $v_F^{(s)}$. In this case we obtain:
 	\begin{equation}
 		J^{(bulk)}_x =  2 \frac{e(\mu - \mu_0)\beta}{4\pi^2\hbar^2 }\,   \frac{E/B_\bot}{v_F^{(s)}}\frac{L}{1+\frac{v_F}{v_F^{(s)}}\frac{\beta/(e B_\bot)}{H}}\label{J_0}
 	\end{equation}
 In addition we consider separately the quasiparticles living on the upper and the lower sides of the sample. The correponding two - dimensional density  
(for each pair of Weyl points) is $2 n_s = 2 \frac{(\mu - \mu_0) \beta}{(2 \pi \hbar)^2 v_F^{(s)}}$. 
The resulting contribution to electric current is 
 \begin{equation}
 	J^{(boundary)}_x =2 \frac{ e (\mu - \mu_0) \beta}{4\pi^2\hbar^2 } \, \frac{ \frac{1}{1 - \frac{E/B_\bot}{v_F^{(s)}}} - \frac{1}{1 + \frac{E/B_\bot}{v_F^{(s)}}} }{\frac{1}{1 - \frac{E/B_\bot}{v_F^{(s)}}} + \frac{1}{1 + \frac{E/B_\bot}{v_F^{(s)}}} + 2\frac{v_F^{(s)}e H B_\bot}{v_F \beta}  }L
 \end{equation}
For small enough values of $E$ we obtain:  
\begin{equation}
	J^{(boundary)}_x = 2 \frac{e(\mu - \mu_0)\beta}{4\pi^2\hbar^2 }\,   \frac{E/B_\bot}{v_F^{(s)}}\frac{L}{1+\frac{v^{(s)}_F}{v_F}\frac{H}{\beta/(e B_\bot)}}\label{J_1}
\end{equation}
In total we obtain: 
\begin{eqnarray}
	J_x &=& 2 \frac{e(\mu - \mu_0)\beta}{4\pi^2\hbar^2 }\,   \frac{E/B_\bot}{v_F^{(s)}}L \Bigl(\frac{1}{1+\frac{v^{(s)}_F}{v_F}\frac{H}{\beta/(e B_\bot)}}\nonumber\\ && + \frac{1}{1+\frac{v_F}{v_F^{(s)}}\frac{\beta/(e B_\bot)}{H}} \Bigr)\label{J_12}
\end{eqnarray}
 This expression is simplified considerably:
 \begin{eqnarray}
 	J_x &=& 2 \frac{e(\mu - \mu_0)\beta}{4\pi^2\hbar^2 }\,   \frac{E/B_\bot}{v_F^{(s)}}L \label{J_123}
 \end{eqnarray}
  Again, we made this estimate for one pair of Weyl points. In case of the sample with several pairs the answer for Hall current is not changed. Then instead of the distance $\beta$ between the two Weyl points of opposite chirality, we substitute into Eq. (\ref{J_123}) the absolute value $|\vec{\beta}|$ of vector $\vec{\beta} = \sum_k \vec{\beta}_k$.  

 Then the corresponding contribution to the Hall conductance is
\begin{equation}
	\Sigma^{Weyl}_{xy} = 2\frac{e(\mu-\mu_0)\beta}{4\pi^2\hbar^2 }\,  \frac{1}{B_\bot v_F^{(s)}}L \label{Sigma_}
\end{equation}
Notice that in this expression Fermi velocity $v_F^{(s)}$ is defined along the Fermi arc. 
 The value of $\Sigma^{Weyl}_{xy}$ is to be compared with $\Sigma^{(0)}_{xy}$:
\begin{equation}
	\frac{\Sigma^{Weyl}_{xy}}{\Sigma^{(0)}_{xy}} = \frac{l}{{\cal N} H}, \label{SigmaWSigma0}
\end{equation}
where $l = \beta/(e B_\bot)$ is the average length of the part of the Weyl orbit that resides on the upper (or lower) side of the sample. Thus we are interested in the situation, when hight $H$ of the sample  is so small that it becomes compared to $l$. If $H$ becomes much smaller than $l$, then the contribution of Weyl orbits to Hall conductance dominates. We expect that in this case also the contributions of scattering processes remain sub - dominant. For Co$_3$Sn$_2$S$_2$ the latter condition means
$$
H \ll 0.1\,  mm
$$   
Say, for the sample  with $R = L = 5$ mm, $H = 5$ $\mu$m, $B_\bot = 0.1 $ T and sufficiently small voltage we estimate  
$$
\Sigma^{Weyl}_{xy} \approx 0.1\, \Omega^{-1} cm
$$
while $\Sigma^{(0)} \approx 0.03\Omega^{-1}cm$. 

\begin{figure}[h]
	\centering  %
	\includegraphics[width=0.9\linewidth]{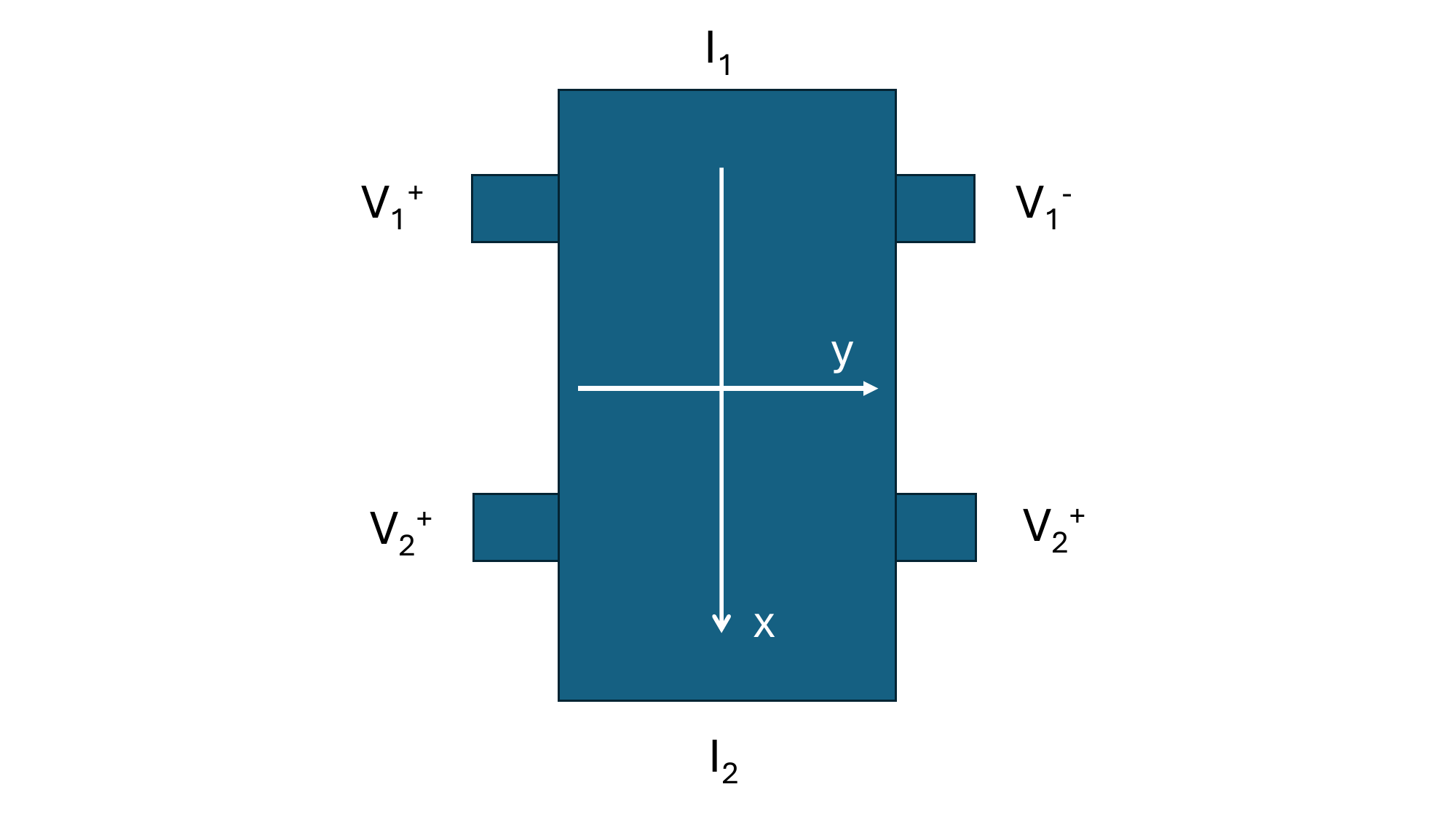}  %
	\caption{We represent here the six - contact Hall bar situated in the $xy$ plane. Vector  $\vec{\beta}$ of the sample is assumed to be directed along axis $y$. Electric current is applied between the contacts $I_1$ and $I_2$ and is to be directed along axis $x$. Axis $z$ is orthogonal to the plane $xy$. The size of the sample in this direction is assumed to be on the order of several $\mu m$ or smaller, while the size along the direction of axes $x$ and $y$ might be on the order of several $mm$. The applied magnetic field $\vec{B}$ might be on the order of $0.1$ T or larger and belongs to the plane $zy$, angle $\theta$ between $\vec{B}$ and axis $z$ belongs to the interval between $0$ and $\pi/2$ and is to be variable. Hall voltage might be measured between the contacts $V_1^+$ and $V_1^-$ or between $V_2^+$ and $V_2^-$. }  %
	\label{HallBar}   %
\end{figure}

For completeness let us further improve our estimate of $\Sigma_{xy}$ taking into account the ordinary resistivity $\rho_\parallel$. (Scattering contributions to Hall resistivity are still neglected.) Then we have the following relation between electric field and electric current density:
\begin{equation}
	{E}_y =  R_1 {M}_\bot {j}_x + \rho_{\parallel} {j}_y  \label{ohm3}
\end{equation} 
Neglecting for simplicity magnetic field $\vec{H}$ inside the sample, we relate the coefficient $R_1$ with the above calculated values of $\Sigma^{Weyl}_{xy}$ and $\Sigma^{(0)}$ (both contributions are summed up in expression for resistivity):
\begin{eqnarray}
R_1  &=& \frac{ HL}{(\Sigma^{Weyl}_{xy}+\Sigma^{(0)}_{xy})M_\bot}
\end{eqnarray}
 Then the value of resulting Hall conductance is  given by
\begin{eqnarray}
	\Sigma_{xy} &=& \frac{R_1 {M}_\bot }{(R_1 M_\bot )^2  + \rho_\parallel^2}LH
	\label{sigmaR2}	
\end{eqnarray}
As the example of Co$_3$Sn$_2$S$_2$  shows, for sufficiently small hight of the sample, when the contribution of Weyl orbits dominates over $\Sigma^{(0)}_{xy}$, the value of $R_1 M_\bot $ remains also much larger than $\rho_\parallel$, and the given improvement does not change much the estimate of Hall conductance. 

The possible experimental setup for the observation of the Hall conductance based on the simplest $6$ - contact Hall bar is represented in Fig. \ref{HallBar}. We propose to use the six - contact Hall bar situated in the $xy$ plane. Vector  $\vec{\beta}$ of the sample is assumed to be directed along axis $y$. Electric current is applied between the contacts $I_1$ and $I_2$ and is to be directed along axis $x$. Axis $z$ is orthogonal to the plane $xy$. The size of the sample in this direction is assumed to be on the order of several $\mu m$ or smaller, while the size along the direction of axes $x$ and $y$ might be on the order of several $mm$. The applied magnetic field $\vec{B}$ might be on the order of $0.1$ T or larger, and belongs to the plane $zy$. Angle $\theta$ between $\vec{B}$ and axis $z$ belongs to the interval between $0$ and $\pi/2$ and is to be variable. Larger values of magnetic field might be required in order to make the CSE effect more pronounced. Then the size of the system in $z$ direction is to be made smaller according to Eq. (\ref{SigmaWSigma0}) so that the Weyl orbit contribution to conductance remains dominating. (Recall that in \cite{moll2016transport}  the samples of width on the order of $0.1 \mu m$ were used.)  Temperature is assumed to be sufficiently small, presumably, of the order of $1$ K or smaller. This is required in order to suppress the temperature corrections to topological effects.  Besides, for the same reason the prepared sample is to be sufficiently clean, i.e. the concentration of impurities is to be minimized during the sample growth.  Hall voltage might be measured between the contacts $V_1^+$ and $V_1^-$ or between $V_2^+$ and $V_2^-$.

\section{Conclusions}

\label{concl}

In the present paper we reviewed application of Wigner - Weyl calculus to calculation of topological quantities in type I Weyl semimetals. We demonstrate that in magnetic Weyl semimetals the intrinsic contribution to anomalous Hall effect and intrinsic contribution to magnetic moment both are expressed through the same quantity
\begin{widetext}
	\begin{eqnarray}
		{\cal M}_i &=&   \frac{1}{3! \, 4\pi^2\,  V} \epsilon_{ijkl} \int d^3 r\, {d^4 p}\, {\rm tr}\,{\bf G}_W^{(0)} \star \partial_{p_j} {\bf Q}_W^{(0)} \star {\bf G}_W^{(0)} \star \partial_{p_k} {\bf Q}_W^{(0)} \star {\bf G}_W^{(0)} \star \partial_{p_l} {\bf Q}^{(0)}_W \label{rez1}
	\end{eqnarray}
\end{widetext}
This quantity would become a topological invariant for topological insulator. In the case of Weyl semimetal it remains robust to those modifications of the system, which do not alter the positions of singularities in phase space of (Wigner transformed) Green function.

It is worth mentioning that the Green function entering Eq. (\ref{rez1}) contains all perturbative contributions. In the non - marginal cases the value of Eq. (\ref{rez1}) may be easily calculated, and it appears to be related intimately to the topological invariant
	\begin{widetext}
	\begin{eqnarray}
		\mathcal{N}_3
		&=&\frac{1}{24 \pi^2 {V}}
		\int d^3x \int_{\Sigma(x)}
		\tr \Bigg[
		{\bf G}_W^{(0)}\star d {\bf Q}_W^{(0)} \star {\bf G}_W^{(0)}
		\wedge \star d {\bf Q}_W^{(0)}\star {\bf G}_W^{(0)} \star \wedge d {\bf Q}_W^{(0)}
		\Bigg]
	\end{eqnarray}
\end{widetext}
protecting Weyl points. Here the integration is over the hypersurface $\Sigma(x)$ that surrounds the given Weyl point. For the minimal values of this invariant we come to the simple expression
\begin{equation}
		\vec{\cal M} =- \vec{\beta} =- \sum_{k} \vec{\beta}_k
\end{equation}
where $\vec{\beta}_k$ is vector connecting Weyl points in the $k$ - th pair of Weyl points. (It points from the left - handed Weyl point to the right - handed Weyl point.)

Chiral separation effect appears to be related to the topological invariant
	\begin{widetext}
	\begin{eqnarray}
		\mathcal{N}
		&=&\frac{1}{48 \pi^2 {V}}
		\int d^3x \int_{\Sigma(x)}
		\tr \Bigg[\gamma^5
		{\bf G}_W^{(0)}\star d {\bf Q}_W^{(0)} \star {\bf G}_W^{(0)}
		\wedge \star d {\bf Q}_W^{(0)}\star {\bf G}_W^{(0)} \star \wedge d {\bf Q}_W^{(0)}
		\Bigg]
	\end{eqnarray}
\end{widetext}
which is expressed as well through the interacting Green functions. One can see that it differs from ${\cal N}_3$ by the factor $1/2$, by the hypersurface $\Sigma(x)$ (it surrounds now all singularities of the Green function $G_W$), and by the presence of factor $\gamma_5(p,x)$ (it is the chirality operator defined in phase space). The natural assumption is that for the interacting system function $G_W(p,x)$ is connected to the one of the non - interacting system by smooth deformation. {\it This way we prove (to the best of our knowledge, for the first time) that the CSE conductivity in the Weyl semimetal is given by its naive expression calculated for the non - interacting system: it is not renormalized by interactions although the latter are strong.}

Using the above mentioned results we develop the theory of Weyl orbits. Originally it was proposed in \cite{Potter_2014} that these orbits consist of the bulk motion in coordinate space (which is nothing but the CSE current), and sliding along the Fermi arcs along the boundary of the sample. The presence of these Weyl orbits reflects existence of chiral separation effect. Therefore, their detection may serve as the indirect way to observe the CSE experimentally. In \cite{Potter_2014} it was proposed that the semiclassical quantization of Weyl orbits gives discrete values of energies, which may be observed in experiment through frequencies of quantum oscillations.

We propose the alternative (and, possibly, more direct) way to detect existence of Weyl orbits. Nontrivial distribution of electric potential is maintained on the surface of the sample as demonstrated schematically in Fig. \ref{fig5}. Magnetic induction field belongs to plane ($xy$) and is tilted with respect to vertical axis $z$, the total magnetic moment is directed along it, but the intrinsic component of magnetic moment (caused by the difference in the positions of Weyl points of opposite chirality) is directed along horizontal axis $y$.  Voltage in direction of axis $y$ modifies the form of the Weyl orbits, which become open and provide overall drift of electrons in direction of axis $x$. As a result the new contribution to anomalous Hall current appears. The effect is to be seen through the contribution to Hall conductance: 
\begin{equation}
	\Sigma^{Weyl}_{xy} = \frac{e(\mu-\mu_0)\beta}{4\pi^2\hbar^2 }\, L\,  \frac{1}{B_\bot v_F^{(s)}}
\end{equation}
Here $\mu$ is chemical potential, $\mu_0$ is the level of Fermi points, $\beta$ is the absolute value of vector $\vec{\beta}$, $B_\bot$ is projection of magnetic field to axis $z$, $L$ is the size of the sample in $y$ - direction, $v_F^{(s)}$ is Fermi velocity on the surface of the sample along the Fermi arc.  

It is expected that this contribution becomes dominating for the samples of very small sizes in $z$ direction (of the order of several $\mu$m).  
Practical experimental setup, as usual for the experimental investigation of Hall effect assumes application of electric current in direction of axis $x$ and measuring of the Hall voltage along axis $y$. It is worth mentioning that the conventional  intrinsic contribution to  anomalous Hall effect (caused by the difference in positions of Weyl points, see Sect. \ref{AQHE}) is to be seen through measurement of Hall voltage in another direction (the direction of axis $z$).

Thus, we propose the new way to detect existence of Weyl orbits, which provides possibility to discover experimentally the chiral separation effect (together with the previous proposition of \cite{Potter_2014} to measure quantum oscillations in density of states).

Strictly speaking the results presented in the present paper refer to the limit of vanishing temperature. This is because all effects related to topology exist in their pure form at vanishing temperature only. At finite temperatures the corrections to the topological quantities appear. In practise the finite temperature studies are to be performed, but the temperature is assumed to be sufficiently small so that the mentioned finite temperature corrections remain small. The estimate of the maximal possible values of temperature remains far out of the scope of the present paper since it requires the use of more refined techniques and remains highly material – dependent. We expect that in practise the temperatures on the order of several Kelvin or less are needed in order to perform the proposed measurements of the contribution of Weyl orbits to Hall effect. This is the domain of temperatures used in \cite{zhang2017evolution,zheng2017recognition,zhang2019ultrahigh,zhang2019quantum} in the experimental studies of quantum oscillations intended at the observation of Weyl orbits.

\begin{appendices}

	\begin{section}{Euclidean conventions}
	\label{conventions}	
	In the present paper we use the following conventions of vector indices. In relativistic theory there are two types of indices: upper and lower. For vector potential of electromagnetic field $A$ these two indices are related as
	$$
	A^\mu = g^{\mu \nu} A_\nu ,\quad \mu,\nu = 0,1,2,3
	$$  	
	where $g^{\mu\nu} = {\rm diag}(1,-1,-1,-1)$. 
	Correspondingly, lowering of spatial Minkowski space indices is followed by the change of sign, while lowering of the $0$-th component of vector is not followed by change of sign.  
	
	In Euclidean space lowering and lifting of indices does not result in any change of the correspondig components of vector:
	$$
	A^i = A_i, \quad i = 1,2,3,4
	$$
	
	According to our conventions Euclidean components of vector are related to its Minkoswkian components as follows:
	$$
	A^i = A_i = A^\mu = - A_\mu, \quad \mu = i = 1,2,3
	$$
	but
	$$
	A^4 = A_4 = -i A^0 = -i A_0
	$$
	In spite of relation between components of coordinates
	$$
	x^4=x_4 = + i x^0 = + i x_0
	$$
	This is because Matsubara frequency $\omega = p^4$ is related to energy ${\cal E} = p^0$ as $p_4 = \omega = - i {\cal E} = -i p_0$. With this identification we obtain for bare $Q$:
	$$
	Q(\omega,\vec{p}) = i\omega - {\cal H}(\vec{p})
	$$ 
	with the plus sign in front of $i \omega$. Then 
	$$
	Q(p_4 - A_4,\vec{p} - \vec{A}) = i\omega - A_0 - {\cal H}(\vec{p}-\vec{A})
	$$
	as it should.

	In particular, the above conventions are valid for components of the four - vector of electromagnetic potential.
	
	For the field strength $F$ we have:
	$$
	F_{\mu\nu} = \partial_\mu A_\nu - \partial_\nu A_\mu = \frac{\partial}{\partial x^\mu} A_\nu - \frac{\partial}{\partial x^\nu} A_\mu 
	$$
	We define Euclidean field strength using the similar expression:
	$$
	F^{ij} = F_{ij} = - F_{\mu\nu} = \frac{\partial}{\partial x^i} A_j - \frac{\partial}{\partial x^j} A_i, $$$$ \mu=i = 1,2,3, \nu = j=1,2,3 
	$$ 
	Correspondingly, $F_{ij} = \epsilon_{ijk} B_k$, where $\vec{B}$ is magnetic field.
	
	Mixed spatial - time components are:
	$$
	F^{4j} = F_{4j} =\frac{\partial}{\partial x^4} A_j -  \frac{\partial}{\partial x^j} A_4 $$$$=i \partial_{x^0} A_\nu + i \partial_{x^\nu} A_0, \quad \nu = j=1,2,3 
	$$ 
	and
	$F_{4j} = -i E_j$ since we do not consider vector potential depending on time.

	\end{section}

\section{Wigner Weyl calculus applied to magnetic Weyl semimetals} 
\label{AppendixB}
	
\subsection{Iterative solution of Groenewold equation}

Renormalized "Dirac operator" and renormalized electron Green function obey
\be \hat {\bf Q} \hat {\bf G}=1 \ee
After Wigner transformation this equation becomes the so - called  Groenewold equation
\be {\bf Q}_W(p,x) \star  {\bf G}_W(p,x)=1 \label{Groen}\ee

It is assumed that external fields (electromagnetic field, field of impurities, elastic deformations, etc) vary slowly: they may be neglected at the distance of the order of the  lattice cell size. Under this supposition Weyl symbol of non - interacting Dirac operator depends on external electromagnetic field $A$ as  ${ Q}_W(p,x)={ Q}^{(0)}_W(p-A(x),x)$. The explicit dependence on coordinates results here from the other sources of inhomogeneity.
At the same time ${\bf Q}_W(p,x)$  can be represented as
\begin{equation}
	{\bf Q}_W(p,x) = {\bf Q}^{(0)}_W(p-A(x),x) + {\bf Q}^{(1)}_{(ij) \,W}(p-A(x),x)F_{ij} + ...\label{QpA}
\end{equation}
where $F_{ij}$ is external electromagnetic field strength. Dots stand for the terms proportional to the higher powers of $F$ and the derivatives of $F$. This expansion is valid for sufficiently weak external field if $|\lambda^2 F_{ij}| \ll 1$, where $\lambda$ is the typical correlation length. Notice that the correlation length associated with bare Dirac operator is equal to the lattice cell size. When  interactions are turned on, the correlation length becomes much larger.

We will use expansion  \cite{ZW2019}
$$
\star = 1+\frac{i}{2} \left( \overleftarrow{\partial}_{x}\overrightarrow{\partial_p}-\overleftarrow{\partial_p}\overrightarrow{\partial}_{x}\right )+...
$$
and represent  the solution of Groenewold equation (\ref{Groen}) (up to the terms linear in $F$) as
\begin{equation}
	{\bf G}_W(p,x) \approx {\bf G}^{(0)}_W(p,x)+{\bf G}_{(ij)W}^{(1)}F_{ij} \label{exp}
\end{equation}
Here  ${\bf G}^{(0)}_W(p,x)$ is solution of reduced Groenewold equation (where $A(x)$ is omitted):
$$
{\bf G}^{(0)}_W(p,x)\star {\bf Q}^{(0)}_W(p,x)=1
$$
The first order term in derivative of $A$ has the form
\begin{eqnarray}
	{\bf G}_{(ij)W}^{(1)}&=&
	\frac{i}{2} \Bigl[ {\bf G}_W^{(0)}\star (\pd_{p_i} {\bf Q}_W^{(0)}) \star {\bf G}_W^{(0)}
	\star (\pd_{p_j} {\bf Q}_W^{(0)}) \star {\bf G}_W^{(0)} \Bigr]\nonumber\\ && - {\bf G}_W^{(0)}\star {\bf Q}_{(ij)W}^{(1)} \star {\bf G}_W^{(0)} \label{exp1}
\end{eqnarray}
Notice that we use here the Euclidean conventions of indexes for vectors and tensors (see Appendix \ref{conventions}).

\subsection{QHE conductivity}
\label{QHE}
The local renormalized electric current density can be taken in the form
\be
{\bf j}_k(x)= -
\int_{\mathcal M} \frac{d^Dp}{(2\pi)^D}
\tr \left[  {\bf G}_W(x,p) \pd_{p_k} {\bf Q}_W(x,p)  \right]
\label{jix}\ee
The linear response to external field strength gives:
\be
{\bf j}_k(x)&=&-
\frac{i}{2}
\int_{\mathcal M} \frac{d^Dp}{(2\pi)^D}
\tr \Bigl[
\Big[ {\bf G}_W^{(0)}\star (\pd_{p_i} {\bf Q}_W^{(0)})\nonumber \\&&\star {\bf G}_W^{(0)}
\star (\pd_{p_j} {\bf Q}_W^{(0)}) \star {\bf G}_W^{(0)} \Big] \pd_{p_k}{\bf Q}_W^{(0)}
\Bigr]
F_{ij}\nonumber\\
&& + \int_{\mathcal M} \frac{d^Dp}{(2\pi)^D}
\tr \Bigl[\Big[ {\bf G}_W^{(0)}\star {\bf Q}_{(ij)W}^{(1)} \star {\bf G}_W^{(0)}  \Big] \pd_{p_k}{\bf Q}_W^{(0)}
\Bigr]
F_{ij}
\nonumber\\
&& - \int_{\mathcal M} \frac{d^Dp}{(2\pi)^D}
\tr \Bigl[{\bf G}_W^{(0)}   \pd_{p_k} \Big[{\bf Q}_{(ij)W}^{(1)}\Big]
\Bigr]
F_{ij}
\label{jilr}\ee
Electric current averaged over the whole system volume $V$ is
\be
&&\bar{J}_i\equiv \frac{1}{\beta {V}}\sum_x {\bf j}_i(x)\nonumber\\&&=
-\frac{1}{\beta {v}}\int d^Dx \int_{\mathcal M} \frac{d^Dp}{{V}|\mathcal M|}
\tr \left[ {\bf G}_W(x,p) \partial_{p_i} {\bf Q}_W(x,p)  \right]\nonumber\\
&&=-\frac{1}{\beta {V}}\Tr \left[  {\bf G}_W(x,p) \partial_{p_i} {\bf Q}_W(x,p)  \right]
\label{Ii}\ee
Here $v$ is volume of the lattice cell. As above we use formula ${ v}|\mathcal M| = (2\pi)^D$ and obtain
\be
\bar{J}_k&&=
-\frac{i}{2}\frac{1}{\beta{V}}
\int d^Dx\int_{\mathcal M} \frac{d^Dp}{(2\pi)^D}\tr \Bigl[
\Bigl[ {\bf G}_W^{(0)}\star (\pd_{p_i} {\bf Q}_W^{(0)})\nonumber \\&&\star {\bf G}_W^{(0)}
\star (\pd_{p_j} {\bf Q}_W^{(0)}) \star {\bf G}_W^{(0)}\pd_{p_k}{\bf Q}_W^{(0)} \nonumber\\&&
+ 2i {\bf G}_W^{(0)}\star {\bf Q}_{(ij)W}^{(1)} \star {\bf G}_W^{(0)} \pd_{p_k}{\bf Q}_W^{(0)}
\nonumber\\&& - 2i  {\bf G}_W^{(0)}   \pd_{p_k} \Big[{\bf Q}_{(ij)W}^{(1)}\Big] \Bigr]
\Bigr]
F_{ij}
\label{Jilr}\ee
The last two terms in this expression cancel each other, and we arrive at
\be
\bar{J}_k&&=
-\frac{i}{2}\frac{1}{\beta{V}}
\int d^Dx\int_{\mathcal M} \frac{d^Dp}{(2\pi)^D}\tr
\Bigl[ {\bf G}_W^{(0)}\star (\pd_{p_i} {\bf Q}_W^{(0)})\nonumber \\&&\star {\bf G}_W^{(0)}
\star (\pd_{p_j} {\bf Q}_W^{(0)}) \star {\bf G}_W^{(0)}\pd_{p_k}{\bf Q}_W^{(0)}
\Bigr]
F_{ij}
\label{Jilr2}\ee
We take into account that the Euclidean field strength differs by the factor $-i$ from the ordinary one. This gives the following expression for the Hall conductivity \cite{ZW2019}:
\begin{equation}
	\sigma^H_{ij} = \frac{e^2}{h}\, \frac{\epsilon_{ijk}{\cal M}_k}{2\pi}\label{sigmaH}
\end{equation}
with
\begin{eqnarray}
	{\cal M}_i &=&   \frac{1}{3! \, 4\pi^2\,  V} \epsilon_{ijkl} \int d^3 r\, {d^4 p}\, {\rm tr}\,{\bf G}_W^{(0)} \star \partial_{p_j} {\bf Q}_W^{(0)} \nonumber\\&& \star {\bf G}_W^{(0)} \star \partial_{p_k} {\bf Q}_W^{(0)} \star {\bf G}_W^{(0)} \star \partial_{p_l} {\bf Q}^{(0)}_W
\end{eqnarray}

\subsection{Response of axial current to magnetic field}

In a similar way we can consider the renormalized axial current density
\be
{\bf j}^5_k(x)= -
\int_{\mathcal M} \frac{d^Dp}{(2\pi)^D}
\tr \left[ \gamma^5 {\bf G}_W(x,p) \pd_{p_k} {\bf Q}_W(x,p)  \right]
\label{ji5x}\ee
Its response to external field strength is
\be
{\bf j}_k^5(x)&=&-
\frac{i}{2}
\int_{\mathcal M} \frac{d^Dp}{(2\pi)^D}
\tr \Bigl[\gamma^5
\Big[ {\bf G}_W^{(0)}\star (\pd_{p_i} {\bf Q}_W^{(0)})\nonumber \\&&\star {\bf G}_W^{(0)}
\star (\pd_{p_j} {\bf Q}_W^{(0)}) \star {\bf G}_W^{(0)} \Big] \pd_{p_k}{\bf Q}_W^{(0)}
\Bigr]
F_{ij}\nonumber\\
&& + \int_{\mathcal M} \frac{d^Dp}{(2\pi)^D}
\tr \Bigl[\gamma^5
\Big[ {\bf G}_W^{(0)}\star {\bf Q}_{(ij)W}^{(1)} \star {\bf G}_W^{(0)}  \Big] \pd_{p_k}{\bf Q}_W^{(0)}
\Bigr]
F_{ij}
\nonumber\\
&& - \int_{\mathcal M} \frac{d^Dp}{(2\pi)^D}
\tr \Bigl[\gamma^5 {\bf G}_W^{(0)}   \pd_{p_k} \Big[{\bf Q}_{(ij)W}^{(1)}\Big]
\Bigr]
F_{ij}
\label{ji5lr}\ee
Average value of local current over the whole system volume is
\be
&&\bar{J}_i^5\equiv \frac{1}{\beta {V}}\sum_x {\bf j}_i^5(x)\nonumber\\&&=
-\frac{1}{\beta {V}}\int d^Dx \int_{\mathcal M} \frac{d^Dp}{{ v}|\mathcal M|}
\tr \left[\gamma^5  {\bf G}_W(x,p) \partial_{p_i} {\bf Q}_W(x,p)  \right]\nonumber\\
&&=-\frac{1}{\beta {V}}\Tr \left[ \gamma^5 {\bf G}_W(x,p) \partial_{p_i} {\bf Q}_W(x,p)  \right]
\label{Ii5}\ee
that is
\be
\bar{J}_k^5&&=
-\frac{i}{2}\frac{1}{\beta{V}}
\int d^Dx\int_{\mathcal M} \frac{d^Dp}{(2\pi)^D}\tr \Bigl[\gamma^5
\Bigl[ {\bf G}_W^{(0)}\star (\pd_{p_i} {\bf Q}_W^{(0)})\nonumber \\&&\star {\bf G}_W^{(0)}
\star (\pd_{p_j} {\bf Q}_W^{(0)}) \star {\bf G}_W^{(0)}\pd_{p_k}{\bf Q}_W^{(0)} \nonumber \\&&
+ 2i {\bf G}_W^{(0)}\star {\bf Q}_{(ij)W}^{(1)} \star {\bf G}_W^{(0)} \pd_{p_k}{\bf Q}_W^{(0)}
\nonumber \\&& - 2i  {\bf G}_W^{(0)}   \pd_{p_k} \Big[{\bf Q}_{(ij)W}^{(1)}\Big] \Bigr]
\Bigr]
F_{ij}
\label{Ji5lr}\ee
The last two terms do not cancel here each other unlike the case of electric current because $\gamma_5(p,x)$ depends on momentum, and affects integration by parts.


\subsection{Axial current at finite temperature}

Let us consider the expression for the axial current at finite temperature.
Matsubara frequencies are given by
$
p_4=\omega_n=\frac{2\pi(n+\frac{1}{2})}{\beta}
$.
The inverse temperature $\beta = 1/T$ is expressed here in lattice units:
$
N_t\equiv\frac{1}{T}
$. Therefore, the values of $p_4$ are
$
p_4=\frac{2\pi(n_4+\frac{1}{2})}{N_t}$, $ n_4=-\frac{N_t}{2},..,\frac{N_t}{2}-1
$,
which belong to the interval between
$
\om_{n=-\frac{N_t}{2}}=\frac{2\pi(-\frac{N_t}{2}+\frac{1}{2})}{N_t}=
-\pi+\frac{\pi}{N_t}
$ and
$
\om_{n=\frac{N_t}{2}-1}=\frac{2\pi(\frac{N_t}{2}-\frac{1}{2})}{N_t}=
\pi-\frac{\pi}{N_t}
$.
The frequencies that are close to zero are:
$
\om_{n=0}=\frac{\pi}{N_t}
$ and
$
\om_{n=-1}=-\frac{\pi}{N_t}
$.
The values of $\om_n$ are always nonzero. As a result the Green function does not have poles in momentum space.

We obtain the following expression for the axial current
\be
&&\bar{J}_k^5=-
\frac{i}{2}\frac{1}{\beta{V}}
\sum_{n=-\frac{N_t}{2}}^{\frac{N_t}{2}-1}
\int d^3x \int_{\mathcal M_3} \frac{d^3p}{(2\pi)^3}\nonumber\\
&&\tr \Bigl[\gamma^5
\Bigl[ {\bf G}_W^{(0)}\star (\pd_{p_i} {\bf Q}_W^{(0)}) \nonumber\\&&\star {\bf G}_W^{(0)}
\star (\pd_{p_j} {\bf Q}_W^{(0)}) \star {\bf G}_W^{(0)}\pd_{p_k}{\bf Q}_W^{(0)} \nonumber\\&&
+ 2i {\bf G}_W^{(0)}\star {\bf Q}_{(ij)W}^{(1)} \star {\bf G}_W^{(0)} \pd_{p_k}{\bf Q}_W^{(0)}
\nonumber\\&& - 2i  {\bf G}_W^{(0)}   \pd_{p_k} \Big[{\bf Q}_{(ij)W}^{(1)}\Big] \Bigr]
\Bigr]
F_{ij}
\label{Ji5}\ee

In order to introduce chemical potential we shift frequency:  $\omega_n\ra\omega_n-i\mu$. As a result the response of axial current to variation of chemical potential $\delta \mu$ and to external field strength $F_{\mu \nu}$ can be calculated as
\be
&&\bar{J}_k^5=-
\frac{1}{{2V}\beta}
\sum_{n=-\frac{N_t}{2}}^{\frac{N_t}{2}-1}
\int d^3x \int_{\mathcal M_3} \frac{d^3p}{(2\pi)^3}\nonumber\\
&&\pd_{\omega_n}\tr \Bigl[\gamma^5
\Bigl[ {\bf G}_W^{(0)}\star (\pd_{p_i} {\bf Q}_W^{(0)})\nonumber \\&&\star {\bf G}_W^{(0)}
\star (\pd_{p_j} {\bf Q}_W^{(0)}) \star {\bf G}_W^{(0)}\pd_{p_k}{\bf Q}_W^{(0)} \nonumber\\&&
+ 2i {\bf G}_W^{(0)}\star {\bf Q}_{(ij)W}^{(1)} \star {\bf G}_W^{(0)} \pd_{p_k}{\bf Q}_W^{(0)}
\nonumber\\&& - 2i  {\bf G}_W^{(0)}   \pd_{p_k} \Big[{\bf Q}_{(ij)W}^{(1)}\Big] \Bigr]
\Bigr]
F_{ij}\delta\mu
\label{Ji5mu1}\ee
The above expression may be represented as
\be
\bar{J}_k^5(x)=\mathcal{\sigma}_{ijk}F_{ij}\delta\mu
\label{Ji5mu2}\ee
Here
\be
&&\mathcal{\sigma}_{ijk}=
-\frac{1}{{2V}\beta}
\sum_{n=-\frac{N_t}{2}}^{\frac{N_t}{2}-1}
\int d^3x \int_{\mathcal M_3} \frac{d^3p}{(2\pi)^3}\nonumber\\
&&\pd_{\omega_n}\tr \Bigl[\gamma^5
\Bigl[ {\bf G}_W^{(0)}\star (\pd_{p_i} {\bf Q}_W^{(0)})\nonumber \\&&\star {\bf G}_W^{(0)}
\star (\pd_{p_j} {\bf Q}_W^{(0)}) \star {\bf G}_W^{(0)}\pd_{p_k}{\bf Q}_W^{(0)} \nonumber\\&&
+ 2i {\bf G}_W^{(0)}\star {\bf Q}_{(ij)W}^{(1)} \star {\bf G}_W^{(0)} \pd_{p_k}{\bf Q}_W^{(0)}
\nonumber\\&& - 2i  {\bf G}_W^{(0)}   \pd_{p_k} \Big[{\bf Q}_{(ij)W}^{(1)}\Big] \Bigr]
\Bigr]
\label{Nijk5}\ee
is the CSE conductivity. In the presencce of magnetic field $B$ we have $F_{ij} = \epsilon_{ijk} B_k$. Then
$$
\bar{J}_k^5(x)=\mathcal{\sigma}_{ijk}\epsilon_{ijk^\prime} B_{k^\prime}\delta\mu
$$
We represent expression for the CSE conductivity as
\be
\mathcal{\sigma}_{ijk}=&&
\sum_{n=-\frac{N_t}{2}}^{\frac{N_t}{2}-1}
\pd_{\om_n}\mathcal{\sigma}_{ijk}^{(3)}
\label{Nijk5_1}\ee
where
\be
\mathcal{\sigma}_{ijk}^{(3)}&&=-
\frac{1}{{2V}}
\int d^3x \int_{\mathcal M_3} \frac{d^3p}{(2\pi)^3}\nonumber\\
&&\tr \Bigl[\gamma^5
\Bigl[ {\bf G}_W^{(0)}\star (\pd_{p_i} {\bf Q}_W^{(0)}) \nonumber\\&&\star {\bf G}_W^{(0)}
\star (\pd_{p_j} {\bf Q}_W^{(0)}) \star {\bf G}_W^{(0)}\pd_{p_k}{\bf Q}_W^{(0)} \nonumber\\&&
+ 2i {\bf G}_W^{(0)}\star {\bf Q}_{(ij)W}^{(1)} \star {\bf G}_W^{(0)} \pd_{p_k}{\bf Q}_W^{(0)}
\nonumber\\&& - 2i  {\bf G}_W^{(0)}   \pd_{p_k} \Big[{\bf Q}_{(ij)W}^{(1)}\Big] \Bigr]
\Bigr]
\label{N_3}\ee

\subsection{The limit of small temperature }

In the limit of small temperature $T\to 0$, $N_t\to \infty$, $\frac{\pi}{N_t}=\epsilon \to 0$ we replace the sum over Matsubara frequencies by the integral. But the value $\omega = 0$ is to be excluded from integration:
\be
\sum_{n=-\frac{N_t}{2}}^{\frac{N_t}{2}-1} \tab\ra \tab
\frac{\beta}{2\pi}\int_{-\pi+\ep}^{0-\ep}d\omega+ \frac{\beta}{2\pi}\int_{0+\ep}^{\pi-\ep}d\omega
\ee
Then Eq. (\ref{Nijk5}) becomes
\be
\mathcal{\sigma}_{ijk}&&=
\lim_{\ep\ra0}
\int_{-\pi+\ep}^{0-\ep}d\om
\pd_{\om}\mathcal{\sigma}_{ijk}^{(3)}
+
\int_{0+\ep}^{\pi-\ep} d\om
\pd_{\om}\mathcal{\sigma}_{ijk}^{(3)}\nonumber\\
&&=\lim_{\ep\ra0}\Big[
\mathcal{-\sigma}_{ijk}^{(3)}(-\pi+\ep) + \mathcal{\sigma}_{ijk}^{(3)}(0-\ep)\nonumber\\&&-
\mathcal{\sigma}_{ijk}^{(3)}(0+\ep) + \mathcal{\sigma}_{ijk}^{(3)}(\pi-\ep)\Big]
\label{Nijk_int}\ee
using $
\mathcal{\sigma}_{ijk}^{(3)}(-\pi)=\mathcal{\sigma}_{ijk}^{(3)}(\pi)
$, we obtain
\be
&&\mathcal{\sigma}_{ijk}=
\lim_{\epsilon \to 0}[
-\mathcal{\sigma}_{ijk}^{(3)}(0+\ep) +\mathcal{\sigma}_{ijk}^{(3)}(0-\ep) ]
\label{Nijk_N_3}
\ee
where
\begin{widetext}
	\be
	{\sigma}_{ijk}^{(3)}(\om=0\pm \ep)&&=-\frac{1}{{2V}}
	\int d^3x \int_{\mathcal M_3} \frac{d^3p}{(2\pi)^4}
	\tr \Bigl[\gamma^5
	\Bigl[ {\bf G}_W^{(0)}\star (\pd_{p_i} {\bf Q}_W^{(0)}) \star {\bf G}_W^{(0)}
	\star (\pd_{p_j} {\bf Q}_W^{(0)}) \star {\bf G}_W^{(0)}\star \pd_{p_k}{\bf Q}_W^{(0)} \nonumber\\&&
	+ 2i {\bf G}_W^{(0)}\star {\bf Q}_{(ij)W}^{(1)} \star {\bf G}_W^{(0)} \star \pd_{p_k}{\bf Q}_W^{(0)}
	- 2i  {\bf G}_W^{(0)} \star  \pd_{p_k} \Big[{\bf Q}_{(ij)W}^{(1)}\Big] \Bigr]
	\Bigr]\Bigg|_{\om=0\pm \ep}
	\label{N_3_1}\ee
\end{widetext}

In thermal equilibrium Green function  $\bf G$ and Dirac operator $\bf Q$ do not depend on time. The singularities of expression standing in the integral are placed at $\omega = 0$. At finite $\epsilon$ the singularities remain out of the integration region. (Recall that for the homogeneous theory  at $\omega = 0$ the singularities of expressions standing inside the integral coincide with the Fermi surface.)

\subsection{Topological expression for the CSE conductivity}
\label{CSE}
In Eq. (\ref{Nijk_N_3}) inside the integrals the contributions of the two surfaces $\omega = \pm \epsilon$ cancel each other everywhere except for the small region around the singularities. As a result it is enough to perform integration in (\ref{N_3_1}) in these small regions surrounding the   singularities.

Recall that according to our definition  $\gamma^5$ is a factor that depends on position in phase space. It is equal to $+1$ in a small vicinity of right - handed Weyl point (its position, in general, depends on coordinates), and is equal to $-1$ in a small vicinity of left - handed Weyl point, and interpolates between the two in the other points of phase space.

Now the last two terms in Eq. (\ref{N_3_1}) cancel each other because in each relevant region of momentum space $\gamma_5(p,x)$ is either $+1$ or $-1$ depending on the chirality of the Weyl point. The sum of the integrals in Eq. (\ref{Nijk_N_3}) represents a topological invariant. It does not depend on the form of the surface in $4D$ momentum space surrounding the singularities (as long as this surface remains close to the Weyl point). We deform the surface in such a way that it consists of small hyperspheres surrounding the singularities.

Therefore, we obtain
$$\sigma_{ijk} = \epsilon_{ijk} \sigma_{CSE}/2
$$
with
\begin{equation}
	\sigma_{CSE} = \frac{\mathcal{N}}{2\pi^2}\label{sigmaH}
\end{equation}
and
\begin{widetext}
	\begin{eqnarray}
		\mathcal{N}
		&=&\frac{1}{48 \pi^2 {V}}
		\int d^3x \int_{\Sigma(x)}
		\tr \Bigg[\gamma^5
		{\bf G}_W^{(0)}\star d {\bf Q}_W^{(0)} \star {\bf G}_W^{(0)}
		\wedge \star d {\bf Q}_W^{(0)}\star {\bf G}_W^{(0)} \star \wedge d {\bf Q}_W^{(0)}
		\Bigg]
	\end{eqnarray}
\end{widetext}
Here integration in momentum space is over the hypersurface $\Sigma(x)$ that surrounds the positions of the Weyl points.

 Index $^{(0)}$ points out that external electromagnetic field is set to zero, while chemical potential is set to the level of the Weyl points. This level  is assumed to be equal for all Weyl points.

	\begin{section}{Classical consideration of motion in the bulk}
		
	\label{clas}
	
	Let us recall that magnetic induction field $\vec{B}$ belongs to plane ($yz$) and angle between axis $z$ and $\vec{B}$  is $\theta$.  
	The dynamics is considered here in the bulk, and boundary is not  taken into account.	
		Let us denote $\vec{\pi} = \vec{p} - e \vec{A}(r)$. We consider the gauge with $A_y = -Bx\,{\rm cos}\, \theta, A_x = 0, A_z = - Bx\,{\rm sin}\, \theta$. Then close to the Weyl point the classical Hamiltonian in the bulk receives the form
		\begin{equation}
			H(p,r) = v_F |\vec{p}|
		\end{equation}
		We do not take into account anisotropy of Fermi velocity, for simplicity.
		
		It is useful to introduce new coordinates $\tilde{y},\tilde{z}$ instead of $y, z$:
		$$
		\tilde{z} = z \,{\rm cos}\, \theta - y \,{\rm sin}\, \theta, \quad \tilde{y} = z \,{\rm sin}\, \theta + y \,{\rm cos}\, \theta
		$$
		Axis $\tilde{z}$ is directed along magnetic field
		
		We obtain classical action defined in phase space
		\begin{eqnarray}
			S &=& \int \Big( \vec{\pi}d\vec{r} + e B x d \tilde{y} - v_F |\vec{\pi}| dt\Big)\\& = &
			\int \Big( \pi_x d x + \pi_{\tilde{y}} d\tilde{y} + \pi_{\tilde{z}} d\tilde{z} - e B x d \tilde{y} - v_F \sqrt{\pi_x^2 + \pi_{\tilde{y}}^2 + \pi_{\tilde{z}}^2} dt\Big)\nonumber
		\end{eqnarray}
		Classical equations of motion in the bulk (close to Weyl point W$_+$) read
		\begin{eqnarray}
			&&d x = \frac{v_F}{\sqrt{\pi_x^2 + \pi_{\tilde{y}}^2 + \pi_{\tilde{z}}^2}} \pi_x dt\nonumber\\ &&
			d {\tilde{y}} = \frac{v_F}{\sqrt{\pi_x^2 + \pi_{\tilde{y}}^2 + \pi_{\tilde{z}}^2}} \pi_{\tilde{y}} dt\nonumber\\ &&
			d {\tilde{z}} = \frac{v_F}{\sqrt{\pi_x^2 + \pi_{\tilde{y}}^2 + \pi_{\tilde{z}}^2}} \pi_{\tilde{z}} dt\nonumber\\ &&0 = - d\pi_x - eBd{\tilde{y}}\nonumber\\ &&
			0 = -d\pi_{\tilde{y}} + eB dx\nonumber\\&&
			0 = - d\pi_{\tilde{z}}
		\end{eqnarray}
		In particular, it follows that
		$$
		d|\vec{\pi}|^2 = \frac{-eBv_F \pi_x\pi_{\tilde{y}} + eBv_F \pi_{\tilde{y}}\pi_x}{|\vec{\pi}|} dt =0
		$$	
		Let us denote
		$$
		Y = {\tilde{y}} + \frac{\pi_x}{eB}, \quad X = x -  \frac{\pi_{\tilde{y}}}{eB}
		$$
		Solutions of this system of equations reads
		\begin{eqnarray}
			\pi_{\tilde{z}} &=& const \nonumber\\
			{\tilde{z}} & = & const + \frac{v_F}{|\vec{\pi}|}\pi_{\tilde{z}} t\nonumber\\
			|\vec{\pi}| &=& const \nonumber\\
			\pi_x & = & \frac{|\vec{\pi}|}{v_F} \dot{x}\nonumber\\
			\pi_{\tilde{y}} & = & \frac{|\vec{\pi}|}{v_F} \dot{{\tilde{y}}}\nonumber\\
			X & = & x - \frac{p_{\tilde{y}} + eB x}{eB} = -\frac{p_{\tilde{y}}}{eB} = const \nonumber\\
			Y & = & {\tilde{y}} + \frac{p_x }{eB} =  const
		\end{eqnarray}
		we define
		$$
		x = X + \zeta_x, \quad {\tilde{y}} = Y + \zeta_{\tilde{y}}
		$$
		Point in $(x{\tilde{y}})$ plane with coordinates $(X,Y) = (-\frac{p_{\tilde{y}}}{eB}, {\tilde{y}} + \frac{p_x}{eB})$ is called the center of orbit. Then $\vec{\zeta} = (\zeta_x,\zeta_{\tilde{y}})= \frac{1}{eB}(\pi_{\tilde{y}},-\pi_x)$ are the coordinates of the trajectory counted from the center of orbit. For these coordinates we have
		$$
		|\vec{\zeta}|^2 =  \frac{1}{e^2 B^2}(\pi_x^2 + \pi_{\tilde{y}}^2) = r^2 = \frac{1}{e^2 v_F^2 B^2} (\epsilon^2 - v_F^2 \pi_{\tilde{z}}^2)= const
		$$
		which means that the orbit has the form of a circle of constant radius $r$.
		We have
		\begin{eqnarray}
			-e B \zeta_{\tilde{y}} =	\pi_x & = & \frac{|\vec{\pi}|}{v_F} \dot{\zeta_x}  \nonumber\\
			eB \zeta_x =	\pi_{\tilde{y}} & = & \frac{|\vec{\pi}|}{v_F} \dot{\zeta_{\tilde{y}}}\nonumber\\
		\end{eqnarray}
		With parametrization
		$$
		\zeta_x = r \,{\rm cos}\,\phi, \quad \zeta_{\tilde{y}} = r \,{\rm sin}\,\phi
		$$
		we obtain
		\begin{eqnarray}
			-	e B  r \,{\rm sin}\,\phi  & = & -\frac{|\vec{\pi}|}{v_F} r \,{\rm sin}\,\phi \dot{\phi}  \nonumber\\
			e B r \,{\rm cos}\,\phi  & = & \frac{|\vec{\pi}|}{v_F} r \,{\rm  cos}\,\phi \dot{\phi}\nonumber\\
		\end{eqnarray}
		that is
		$$
		\dot{\phi} = \frac{e B v^2_F}{\epsilon} = \omega_B
		$$
		where $\epsilon = v_F |\vec{\pi}|$ is energy of the particle.
		
		For the dependence of polar angle $\phi$ on time we have 	
		\begin{eqnarray}
			\phi = const + \omega_B t
		\end{eqnarray}
	
	The considered motion  contributes the adiabatic invariant entering the semi - classical quantization condition by
	\begin{eqnarray}
		I_h &=& \int \Big(\pi_x d\zeta_x + \pi_{\tilde{y}} d\zeta_{\tilde{y}} + \pi_{\tilde{z}} d {\tilde{z}} + e A_x(x,{\tilde{y}}) dx + e A_{\tilde{y}}(x,{\tilde{y}}) d {\tilde{y}}  \Big)\nonumber\\
		& = &
		\int eB r^2 \,  d\phi  + \int \vec{A}d\vec{r} + \pi_{\tilde{z}} \int d{\tilde{z}} 
	\end{eqnarray}
	For the case when rotation around axis $z$ is absent at all, the first two terms vanish.
		
	\end{section}
	
\end{appendices}

\bibliography{QHE.bib,CSE_MZ.bib,QFTMacroMotion.bib}

\end{document}